\documentclass[12pt]{article}
\usepackage{jheppub} 
\usepackage{amsfonts,amsmath,amssymb,epsf,mathtools}
\usepackage{amsmath,braket}
\usepackage{graphicx,hyperref,color}
\usepackage{pgfplots}
\usepackage{comment}
\hypersetup{
    bookmarks=true,         
    unicode=false,          
    pdftoolbar=true,        
    pdfmenubar=true,        
    linktocpage=true,       
    pdffitwindow=false,     
    pdfstartview={FitH},    
    pdfnewwindow=true,      
    colorlinks=true,       
    linkcolor=blue,          
    citecolor=blue,        
    filecolor=blue,      
    urlcolor=blue           
}

\numberwithin{equation}{section}									

\newcommand{\de}{\partial}
\newcommand{\be}{\begin{equation}}
\newcommand{\ba}{\begin{eqnarray}}
\newcommand{\ea}{\end{eqnarray}}
\newcommand{\ee}{\end{equation}}

\newcommand{\s}{\sqrt}

\newcommand{\ti}{\tilde}
\newcommand{\ap}{\alpha}

\newcommand{\ddd}{\cdot\cdot\cdot}
\newcommand{\no}{\nonumber \\}
\newcommand{\la}{\langle}
\newcommand{\lb}{\rangle}
\newcommand{\bea}{\begin{eqnarray}}
\newcommand{\eea}{\end{eqnarray}}
\newcommand{\bes}{\begin{equation*}}
\newcommand{\beas}{\begin{eqnarray*}}
\newcommand{\eeas}{\end{eqnarray*}}
\newcommand{\bas}{\begin{array*}}
\newcommand{\eas}{\end{array*}}
\newcommand{\ees}{\end{equation*}}
\newcommand{\nn}{\nonumber}

\newcommand{\ep}{\epsilon}

\newcommand{\Tr}{\textrm{Tr}}

\let\a=\alpha         \let\l=\lambda  
  \let\r=\rho 
      
 \let\D=\Delta         
\def\nn{\nonumber}

\usepackage{amsmath}

\subheader{\today}
\title{\boldmath Probing de Sitter Space Using CFT States}

\author[a]{Kazuki Doi,}
\author[a]{Naoki Ogawa,}
\author[a]{Kotaro Shinmyo,}
\author[a]{Yu-ki Suzuki,}
\author[a,b]{Tadashi Takayanagi}
\affiliation[a]{Center for Gravitational Physics and Quantum Information, Yukawa Institute for Theoretical Physics, Kyoto University,\\
	Kitashirakawa Oiwakecho, Sakyo-ku, Kyoto 606-8502, Japan}
\affiliation[b]{Inamori Research Institute for Science,\\
620 Suiginya-cho, Shimogyo-ku,Kyoto 600-8411 Japan}

\abstract{In this paper we construct CFT states describing a putative holographic dual to local excitations in the three-dimensional de Sitter space (dS), called the bulk local states. We find that the conjugation operation in dS$_3/$CFT$_2$ is notably different from that in AdS$_3/$CFT$_2$. This requires us to combine two bulk local states constructed out of different primary states in a CPT-invariant way. 
This analysis explains why Green's functions in the dS Euclidean vacuum cannot simply be obtained from the Wick rotation of those in AdS. We also argue that this characteristic feature explains the emergence of a time coordinate from the dual Euclidean CFT. We show that the information metric for the quantum estimation of bulk coordinate values replicates the de Sitter space metric.}

\begin{document} 

\begin{flushright}
YITP-24-66\\
\end{flushright}
\maketitle
\flushbottom

\clearpage
\section{Introduction}

To understand the principle of holographic duality \cite{tHooft:1993dmi,Susskind:1994vu} from a truly physical viewpoint, we need to explain the emergence of gravitational spacetime coordinates from a lower-dimensional quantum theory. In the AdS/CFT correspondence \cite{Maldacena:1997re,Gubser:1998bc,Witten:1998qj}, the most well-established example of holography, it is argued that a $d$-dimensional conformal field theory (CFT) is equivalent to a gravitational theory in $d+1$-dimensional anti-de Sitter space (AdS), commonly dubbed as AdS$_{d+1}/$CFT$_d$. One of the fundamental questions surrounding this duality is how an observer placed in CFT$_d$ can perceive the effects of also being in a $d+1$-dimensional AdS. In AdS/CFT, we expect a spacelike direction to emerge from CFT due to the holography. This question becomes more non-trivial for holography in de Sitter spacetime, the so-called dS/CFT correspondence \cite{Strominger:2001pn,Witten:2001kn,Maldacena:2002vr}, because we expect a timelike direction to be produced from a Euclidean CFT by the holography. Refer to \cite{Bousso:2001mw,Spradlin:2001nb,Guijosa:2003ze,Ness:2005qj,Harlow:2011ke,Castro:2023bvo} for later developments on general aspects of dS/CFT.

One simple idea on how to detect the dimensions and size of the spacetime in which an observer lives is to consider how much localized excitation can move without energy cost. In AdS/CFT, this was studied in \cite{Miyaji:2015fia,Goto:2016wme,Goto:2017olq} (see also \cite{Nakayama:2015mva,Verlinde:2015qfa,Lewkowycz:2016ukf} for closely related analyses) by analyzing the CFT state dual to a locally excited state in the bulk AdS, which we call a bulk local state. This can be regarded as a quantum state counterpart of the CFT reconstruction of bulk local operators \cite{Hamilton:2005ju,Hamilton:2006az}. This analysis shows that bulk local states provide a family of CFT states whose parameters correspond to the coordinates of bulk AdS. This reproduces the correct two-point function of a scalar field in AdS. Moreover, the AdS metric can be reproduced from the information metric (Bures metric), a quantity computed from bulk local states, by treating the bulk AdS coordinates as parameters in quantum estimation theory.

The main purpose of this paper is to extend this analysis of bulk local states for AdS/CFT to those for holography in three-dimensional de Sitter space dS$_3$ (refer to \cite{Xiao:2014uea,Chatterjee:2015pha,Chatterjee:2016ifv,Bhowmick:2019nso,Sleight:2020obc,Sleight:2021plv} for earlier arguments on bulk local operators in dS/CFT). Since we still do not know the precise formulation of dS/CFT, we only assume that the physics of gravity on dS$_3$ can be equivalently described by a Euclidean CFT$_2$ and respect the geometric correspondence between the isometries of dS$_3$ and the global conformal symmetry of the CFT. We do not assume any details of dS/CFT, e.g. where the dual CFT$_2$ lives in the dS$_3$ spacetime. The main difference between the previous analysis in AdS and the present one for dS is that we expect Lorentzian time to emerge from Euclidean CFT. Since AdS/CFT tells us that a Euclidean holographic CFT$_2$ is dual to gravity in a three-dimensional hyperbolic space H$_3$, we expect Euclidean CFT$_2$ for dS/CFT to be exotic in some ways. For example, in the explicit examples of dS$_4/$CFT$_3$ \cite{Anninos:2011ui,Ng:2012xp} and dS$_3/$CFT$_2$ \cite{Hikida:2021ese,Hikida:2022ltr}, the dual CFTs turn out to be non-unitary.\footnote{These examples fit nicely with the original version of dS/CFT where the Euclidean CFT lives on the spacelike boundaries at the future and past infinity. There are other approaches to holography in dS. One such approach is the construction of dual field theories by deforming AdS/CFT to a finite cutoff scale \cite{Alishahiha:2004md,Dong:2018cuv,Gorbenko:2018oov}. Another is to employ entanglement entropy as a probe to work out how dS/CFT looks, as studied in e.g. \cite{Narayan:2015vda,Sato:2015tta,Miyaji:2015yva,Geng:2021wcq,Doi:2022iyj,Cotler:2023xku,Kawamoto:2023nki}. Yet another approach is to focus on the static patch of dS, as explored in 
e.g. \cite{Susskind:2021esx,Chandrasekaran:2022cip}. Quantum mechanical descriptions of dS$_2$ have also recently been studied in e.g. \cite{Maldacena:2019cbz,Cotler:2019nbi,Susskind:2022bia,Narovlansky:2023lfz}.}
As we will discuss in this paper, we find that a crucial difference between dS$_3/$CFT$_2$ and AdS$_3/$CFT$_2$ lies in the way quantum states are conjugated. This plays an important role when we identify the precise form of the bulk local states.
We will argue that this difference explains the emergence of Lorentzian time from Euclidean CFT$_2$. Moreover this key property helps us to derive the correct two-point function for the Euclidean vacuum in dS$_3$, which is not simply the Wick rotation of that in AdS/CFT.
Specifically, in order to obtain physical states that yield the correct two-point function, we find that it is necessary to take a CPT-invariant linear combination of the two primary states.

This paper is organized as follows: in section \ref{Sec:AdS}, we give a brief review of bulk local states in AdS$_3/$CFT$_2$. In section \ref{Sec:dSBulklocal}, which is the main section of this paper, we present the construction of bulk local operators in dS$_3$, starting from the isometries of dS$_3$. A careful treatment of the conjugation operation characteristic to dS$_3$ allows us to evaluate the two-point function of the bulk local operators in the dual CFT$_2$ and show that this CFT analysis reproduces the known bulk Green's function. We generalize our construction to $\alpha$-states. Furthermore, we evaluate the information metric of bulk local states and reproduce the de Sitter metric.
In section \ref{Sec:dSglobal}, we extend our construction of bulk local states to the entire global patch of dS$_3$. In section \ref{Sec:CD}, we present our conclusion and discuss future problems. In appendix \ref{ap:pauli}, we explain our argument to derive an identity using Pauli matrices. In appendix \ref{ap:waveads3}, we provide a detailed calculation of the scalar field wave function in AdS$_3$. In appendix \ref{ap:scalar}, we derive the scalar field wave function in dS$_3$.
In appendix \ref{appendix_nonhemitian}, we explain how the CFT dual to dS can be seen as a non-Hermitian system.
In appendix \ref{ap:formula}, we compile useful identities of the $\mathfrak{su}(2)$ algebra used in this paper.

\section*{Notations }\label{Notation}
We summarize some of the important notations that appear in this paper for the readers' convenience. Precise definitions of these states will be explained in later sections.
\subsection*{Wick-rotated bulk local states in de Sitter space}
\begin{align}
    \begin{split}
    \ket{\Psi_{\text{dS,W}}^{\Delta_+}(x)}&=e^{(L_0+\tilde{L}_0)t}e^{-i\phi(L_0-\tilde{L}_0)}e^{i\theta J_1}\sum c_k(L_{-1}\tilde{L}_{-1})^k\ket{\Delta_+},\\
\bra{\Psi_{\text{dS,W}}^{\Delta_{\pm}}(x^{\prime})}&=\la \Delta_{\pm}|
\sum_{k=0}^\infty c_k(\Delta_{\pm}) (L_1\ti{L}_1)^k e^{-i\theta J_1}e^{i\phi J_3}e^{-(L_0+\ti{L}_0)t}.
    \end{split}
\end{align}
\subsection*{Conjugations of primary states}
\begin{align}
\begin{split}
  &|\Delta_\pm\lb^\dagger\equiv\la \widehat{\Delta_{\pm}}|,  \quad \bra{\Delta_{\pm}}^\dagger\equiv\ket{\widehat{\Delta_{\pm}}},\\
&\braket{\widehat{\Delta_\pm}|\Delta_\pm}=0,\quad \braket{{\Delta_\pm}|\Delta_\pm}\equiv1.
\end{split}
\end{align}
\subsection*{Conjugations of bulk local states}
\begin{align}
\begin{split}
    \nu_{\pm}\braket{\Psi_{\text{dS,W}}^{\Delta_{\pm}}(x^{\prime}_A)|\Psi_{\text{dS,W}}^{\Delta_{\pm}}(x)}&=\braket{\widehat{\Psi_{\text{dS,W}}^{\Delta_\mp}}(x^{\prime})|\Psi_{\text{dS,W}}^{\Delta_\pm}(x)}=\frac{e^{\mp\mu(\pi-D_{\text{dS}}(x'_A,x))}}{2i\sin D_{\text{dS}}},\\
    \braket{\Psi_{\text{dS,W}}^{\Delta_{\pm}}(x^{\prime})|\Psi_{\text{dS,W}}^{\Delta_{\pm}}(x)}&=\frac{e^{\pm\mu D_{\text{dS}}}}{2i\sin D_{\text{dS}}},\\
    \ket{\Psi_{\text{dS,W}}^{\Delta_+}(x)}^{\dagger}&\equiv\bra{\widehat{\Phi_{\text{dS,W}}^{\Delta_+}}(x)}=\bra{{\Phi_{\text{dS,W}}^{\Delta_-}}(x_A)}.
\end{split}
\end{align}

\section{Bulk local states in AdS$_3$}\label{Sec:AdS}
In this section, we review the construction of bulk local states in AdS space \cite{Miyaji:2015fia,Nakayama:2015mva}. Specifically, we focus on the AdS$_3$ case.
\subsection{Global $\text{SL}(2,\mathbf{R})_{\rm L}\times \text{SL}(2,\mathbf{R})_{\rm R}$ Virasoro generators from AdS$_3$}
Let us consider the AdS$_3$ embedding in Minkowski space:
\be
ds^2=-dX_{-1}^2-dX_0^2+dX_1^2+dX_2^2.
\ee
The AdS spacetime is defined by the hypersurface 
\be
-X_{-1}^2-X_0^2+X_1^2+X_2^2=-1,
\ee
where we set $R_{\rm AdS}=1$.
Particularly, we consider the following embedding
\begin{align}
\begin{split}
    X_{-1}=\cosh\rho\sin \tau,\quad X_0=\cosh\rho\cos \tau,\\
    X_1=\sinh\rho\cos\phi,\quad X_2=\sinh\rho\sin\phi,
\end{split}
\end{align}
which gives the global AdS$_3$
\be
    ds^2=-\cosh^2{\rho} d\tau^2 + d\rho^2 + \sinh^2{\rho} d\phi^2,
    \label{glads}
\ee
where $0\leq\rho<\infty$, $-\infty<\tau<\infty$, and $0\leq\phi<2\pi$. We set the AdS radius $R_{\rm AdS}=1$ in this paper, unless stated otherwise.
The dual CFT can be defined on the cylinder. The isometries of AdS$_3$ form $\text{SL}(2,\mathbf{R})_{\rm L}\times \text{SL}(2,\mathbf{R})_{\rm R}$, generated by the global Virasoro generators $(L_0, L_{\pm 1})$ and $(\tilde{L}_0, \tilde{L}_{\pm 1})$ in two-dimensional CFT. The explicit expression of the generators reads
\begin{align}
\begin{split}
    L_0&=\frac{i}{2}(\de_\tau+\de_\phi),\quad \tilde{L}_0=\frac{i}{2}(\de_\tau-\de_\phi),\\
    L_{\pm1}&=\frac{i}{2}e^{\pm i(\tau+\phi)}\left[\frac{\sinh\rho}{\cosh\rho}\de_\tau+\frac{\cosh\rho}{\sinh\rho}\de_\phi\mp i\de_\rho\right],\\
    \tilde{L}_{\pm1}&=\frac{i}{2}e^{\pm i(\tau-\phi)}\left[\frac{\sinh\rho}{\cosh\rho}\de_\tau-\frac{\cosh\rho}{\sinh\rho}\de_\phi\mp i\de_\rho\right],
\end{split}
\end{align}
obeying
\be
[L_m,L_n]=(m-n)L_{m+n},\quad [\tilde{L}_m,\tilde{L}_n]=(m-n)\tilde{L}_{m+n}.
\ee
The Hermitian conjugate in Lorentzian signature is taken as follows
\be
\quad (L_n)^\dagger=L_{-n},\quad (\tilde{L}_n)^\dagger=\tilde{L}_{-n}.\label{Virasoro}
\ee
We can confirm this conjugation relation by employing the standard inner product of two functions on AdS$_3$: $\la f,g\lb=\int \s{|g|}d\tau d\rho d\phi f^*(\tau,\rho,\phi) g(\tau,\rho,\phi)$, where the volume form is evaluated as $\s{|g|}=\sinh\rho\cosh\rho$, and by performing integration by parts $\de_{x^i}\to -\de_{x^i}$ for $x^i=\tau,\rho,\phi$.

\subsection{Bulk local states in global AdS$_3$}\label{2.2}
\begin{figure}[hhh]
    		\centering
    \includegraphics[width=5cm]{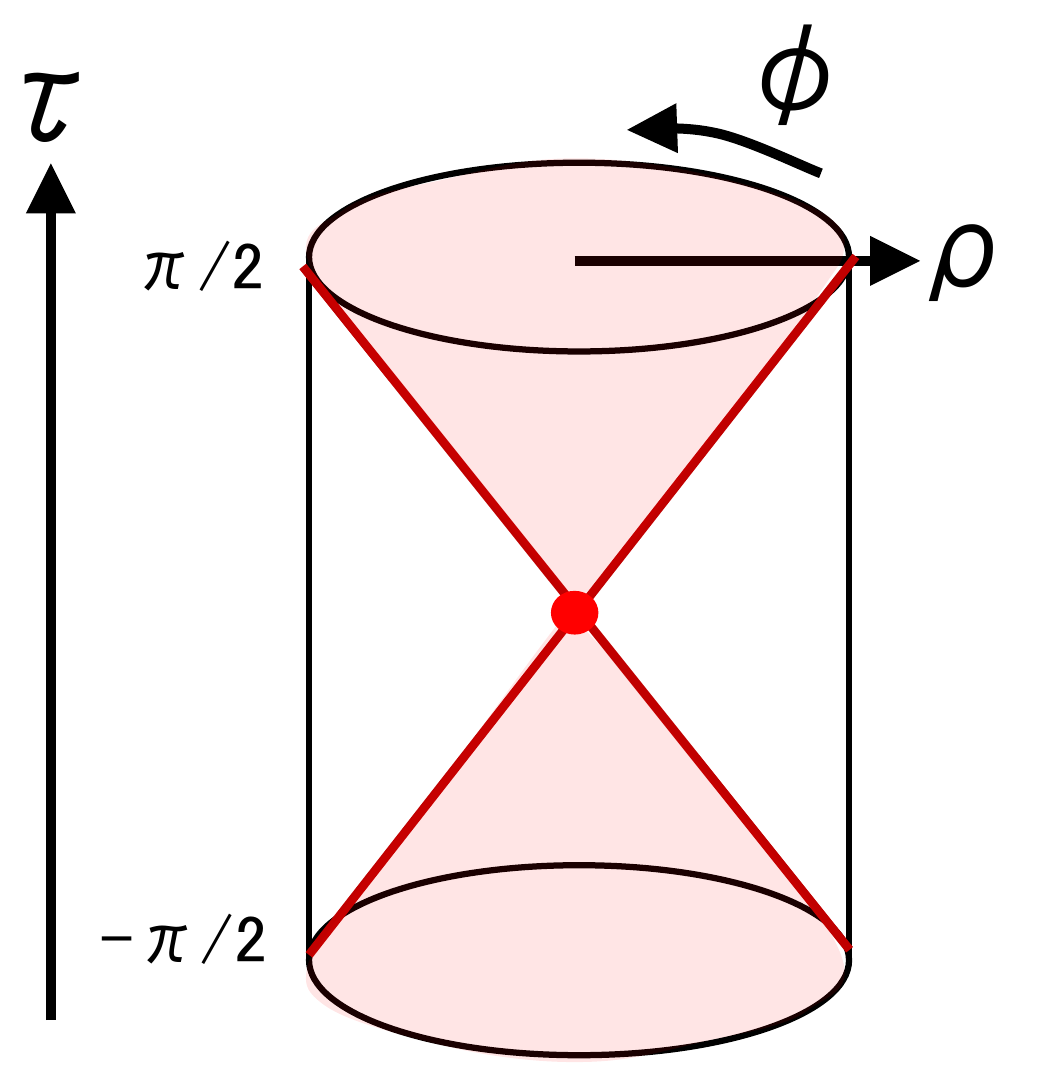}
    		\caption{The light cone $\sinh\rho=|\tan \tau|$ in AdS$_3$, which is invariant under the actions of  
            $L_0-\tilde{L}_0$, $L_1+\tilde{L}_{-1}$, and $L_{-1}+\tilde{L}_1$.} 
    		\label{fig:AdSlocal}
    \end{figure}

Let us first construct a bulk local operator at $\tau=0$, since states at arbitrary $\tau$ can be obtained by time evolution. At $\tau=0$, the generators are defined by
\begin{align}
\begin{split}
    l_0&=L_0-\tilde{L}_0=i\de_\phi,\\
    l_{1}&=\tilde{L}_{-1}-L_1=-ie^{i\phi}\left[\frac{1+\cosh2\rho}{\sinh2\rho}\de_\phi-i\de_\rho\right],\\
    l_{-1}&=\tilde{L}_{1}-L_{-1}=ie^{-i\phi}\left[-\frac{1+\cosh2\rho}{\sinh2\rho}\de_\phi-i\de_\rho\right],
\end{split}
\end{align}
which forms an $\text{SL}(2,\mathbf{R})$ subgroup of $\text{SL}(2,\mathbf{R})_{\rm L}\times \text{SL}(2,\mathbf{R})_{\rm R}$
\be
[l_m,l_n]=(m-n)l_{m+n}.
\ee
Since we can exploit this $\text{SL}(2,\mathbf{R})$ symmetry, suppose that the bulk local state is located at $\tau=\rho=0$, which we shall denote as $\ket{\Psi_{\rm AdS}^{(\Delta)}(\tau=0,\rho=0)}$, where $\Delta$ is the corresponding conformal dimension of the excitation, i.e. the eigenvalue of $L_0+\ti{L}_0$.  The invariance of the point $\tau=\rho=0$ under a certain subgroup of $\text{SL}(2,\mathbf{R})$ forces the constraints
\be
    (L_0-\tilde{L}_0)\ket{\Psi_{\rm AdS}^{(\Delta)}(0,0)}= (L_1+\tilde{L}_{-1})\ket{\Psi_{\rm AdS}^{(\Delta)}(0,0)}= (L_{-1}+\tilde{L}_1)\ket{\Psi_{\rm AdS}^{(\Delta)}(0,0)}=0,
    \label{AdSconst}
\ee
the solution to which is given by
\be
\ket{\Psi_{\rm AdS}^{(\Delta)}(\tau=0,\rho=0)}=e^{\frac{i\pi}{2}(L_0+\tilde{L}_0-\Delta)}\ket{I_{\Delta}},  \label{IshibashiAdS}
\ee
where $\ket{I_{\Delta}}$ is the Ishibashi state for the global Virasoro algebra $\text{SL}(2,\mathbf{R})_{\rm L}\times \text{SL}(2,\mathbf{R})_{\rm R}$, satisfying
\be
(L_n-\tilde{L}_{-n})\ket{I_{\Delta}}=0,\quad n=0,\pm1.  \label{Ishibashic}
\ee
Its explicit form takes
\be
\ket{I_{\Delta}}=\sum_{k=0}^\infty\ket{k}_{\rm L}\ket{k}_{\rm R}.
\label{Ishibashia}
\ee
Here, we introduced the orthonormal descendant states
\be
\ket{k}=\prod_{j=1}^{k}\sqrt{\frac{1}{j^2+(\D-1)j}}(L_{-1})^k\ket{\Delta},\label{Ishibashib}
\ee
such that $\la k_1|k_2 \lb=\delta_{k_1,k_2}$, and the primary state with conformal dimension $\Delta$, defined by
\ba
L_0|\Delta\lb=\tilde{L}_0|\Delta\lb=\frac{\Delta}{2}|\Delta\lb,\ \ \ \ 
L_{1}|\Delta\lb=0.  \label{primarycond}
\ea
We can confirm that the generators 
$L_0-\tilde{L}_0$, $L_1+\tilde{L}_{-1}$ and $L_{-1}+\tilde{L}_1$ in (\ref{AdSconst}) keep the future and past light cones of the point $\tau=\rho=0$ invariant, as depicted in Fig.\ref{fig:AdSlocal}. This is consistent with the time evolution of a locally excited state at the origin at $\tau=0$.

By using the $\text{SL}(2,\mathbf{R})$ transformation and time evolution, we can construct bulk local operators at arbitrary points as
\be
\ket{\Psi_{\rm AdS}^{(\Delta)}(\tau,\rho,\phi)}=e^{-i(L_0+\tilde{L}_0)\tau}e^{i\phi l_0}e^{-\frac{\rho}{2}(l_1-l_{-1})}e^{\frac{i\pi}{2}(L_0+\tilde{L}_0)}\ket{I_{\Delta}}.
\label{bulklop}
\ee
It is straightforward to check that this state satisfies the equation of motion of a scalar field: 
\ba
\left[\frac{1}{2}(L_{-1}L_1+L_1L_{-1})-L_0^2+\frac{1}{2}(\ti{L}_{-1}\ti{L}_1+\ti{L}_1\ti{L}_{-1})-\ti{L}_0^2+\frac{M^2}{2}\right]\ket{\Psi^{\Delta}_{\text{AdS}}(\tau,\rho,\phi)}=0,\no
\ea
where the scalar field mass $M$ is related to the conformal dimension via the standard formula $M^2=\Delta^2-2\Delta$ of AdS$_3/$CFT$_2$.

In this way, we can construct the state (\ref{bulklop}) which describes a local excitation in the bulk AdS$_3$. More explicitly, if we write the bulk scalar field in the bulk quantum field theory as $\hat{\Phi}^{\rm(AdS)}(\tau,\rho,\phi)$, we can simply express the bulk local state as
\ba
\ket{\Psi_{\rm AdS}^{(\Delta)}(\tau,\rho,\phi)}=\hat{\Phi}^{\rm(AdS)}(\tau,\rho,\phi)|0\lb.\label{HKLLS}
\ea
This clarifies the meaning of the bulk local state. We can provide justification for this claim in several ways. First of all, it is invariant under the subgroup of isometries that keep the localized point unchanged, as imposed in (\ref{AdSconst}). Second, as we will see in the next subsection, the inner product $\braket{\Psi_{\rm AdS}^{(\D)}(\tau,\rho,\phi)|\Psi_{\rm AdS}^{(\D)}(0,0,0)}$ becomes divergent at $\tau=\rho=0$. Since this inner product can be thought of as the scalar wave function of the state $|\Psi_{\rm AdS}^{(\D)}(0,0,0)\lb$, this shows that the state is localized at $\tau=\rho=0$.  
Moreover, the inner product agrees with the two-point function of the scalar field $\hat{\Phi}^{\rm(AdS)}(\tau,\rho,\phi)$ in AdS$_3$.
Finally, as shown in \cite{Goto:2016wme}, state $|\Psi_{\rm AdS}^{(\D)}(0,0,0)\lb$ is identical to the state constructed by acting the CFT dual of the bulk local field operator, the so-called HKLL operator \cite{Hamilton:2005ju,Hamilton:2006az}, on the vacuum state, i.e. (\ref{HKLLS}).
It is useful to note that under time evolution, the localized excitation spreads out at the speed of light as confirmed in \cite{Goto:2016wme}, as depicted in Fig.\ref{fig:AdSlocal}.

\subsection{Two-point functions}
As explained in \cite{Miyaji:2015fia}, the scalar wave function in AdS$_3$ can be computed from the inner product between $\ket{\Psi_{\rm AdS}^{(\Delta)}(\tau,\rho,\phi)}$ and each descendant state $|k\lb_L|\bar{k}\lb_R$, which leads to the following form (for $k\geq \bar{k}$)
\be
\Phi^{\rm(AdS)}_{k,\bar{k}}(\tau,\rho,\phi)=e^{-i(\Delta +k+\bar{k})\tau}e^{i(k-\bar{k})\phi}
(\tanh\rho)^{|k-\bar{k}|}(\cosh\rho)^{-\Delta}\cdot P^{(k-\bar{k},\Delta-1)}_{\bar{k}}\left(\frac{2}{\cosh^2\rho}-1\right),\label{wave}
\ee
where the Jacobi polynomial $P$ is defined as 
\ba
P^{(\ap,\beta)}_{n}(x)=\frac{(n+\ap)!}{n!\ap!}\cdot {}_2F_1\left(-n,n+\ap+\beta+1;\ap+1;\frac{1-x}{2}\right).
\ea
For $k<\bar{k}$, in order to obtain the correct expression we simply have to replace $k$ with $\bar{k}$ in the $P$-function. Its derivation can be found in appendix \ref{ap:scalar}.
The Green's function is written as
\begin{align}
    G_{\rm AdS}(\tau,\rho,\phi)&=\braket{\Psi_{\rm AdS}^{(\D)}(\tau,\rho,\phi)|\Psi_{\rm AdS}^{(\D)}(0,0,0)}\nn\\
    &=\bra{\Psi_{\rm AdS}^{(\D)}(\tau,0,0)}e^{-\rho(L_1-L_{-1})}\ket{\Psi_{\rm AdS}^{(\D)}(0,0,0)}.
\end{align}
Using the identity
\be
1=\sum_{k,\bar{k}=0}^\infty\ket{k}_{\rm L}\ket{\bar{k}}_{\rm R}\bra{k}_{\rm L}\bra{\bar{k}}_{\rm R},
\ee
we find that
\be
G_{\rm AdS}(\tau,\rho,\phi)=\sum_{k=0}^\infty(-1)^k\bra{\Psi_{\rm AdS}^{(\D)}(\tau,0,0)}e^{-\rho(L_1-L_{-1})}\ket{k}_{\rm L}\ket{{k}}_{\rm R}.
\ee
After a bit of algebra, we obtain
\be
L_{-1}\tilde{L}_{-1}\Phi^{\rm(AdS)}_{k,k}=-(k+1)(k+\D)\Phi^{\rm(AdS)}_{k+1,k+1}.
\ee
Using this, we can show that
\be
\bra{\Psi_{\rm AdS}^{(\D)}(\tau,0,0)}e^{\rho(L_1-L_{-1})}\ket{k}_{\rm L}\ket{{k}}_{\rm R}=(-1)^k\Phi^{\rm(AdS)}_{k,k}.
\ee
Finally, the Green's function can be evaluated as
\begin{align}
G_{\rm AdS}(\tau,\rho,\phi)&=\sum_{k=0}^\infty\Phi^{\rm(AdS)}_{k,k}(\tau,\rho,\phi)\nn\\
&=\sum_{k=0}^\infty\frac{e^{-i(2k+\D)\tau}}{\cosh^{\D}\rho}P_k^{(0,\D-1)}(1-2\tanh^2\rho)\nn\\
&=\frac{e^{-(\Delta-1)D_{\rm AdS}}}{2\sinh D_{\rm AdS}},\label{AdStwop}
\end{align}
where the geodesic distance $D_{\rm AdS}$ is given by
\be
\cosh D_{\rm AdS}=\cos(\tau-\tau')\cosh\rho\cosh\rho'-\cos(\phi-\phi')\sinh\rho\sinh\rho'.
\ee

\section{Bulk local states in static dS$_3$}\label{Sec:dSBulklocal}
In this section, we construct bulk local states in dS$_3$ by analytically continuing the metric from AdS$_3$.
For the readers' convenience, let us outline the flow of this section. 
In subsection \ref{3.1}, we define the Virasoro generators of the dual CFT$_2$ using the Killing vectors of the dS$_3$ isometry group $\mathrm{SL}(2,\mathbf{C})$.
We then construct bulk local states in static coordinates in subsection \ref{3.2}. 
To discuss the inner product of states constructed in subsection \ref{3.2}, we define a bra state $\bra{\Psi_{\text{dS,W}}^{\Delta_{\pm}}(x^{\prime})}$ in subsection \ref{3.3} through an analytic continuation of the two-point function on AdS spacetime.
We note that this bra state is not defined via the Hermitian conjugate in the dual CFT$_2$.
In subsection \ref{3.5}, we explicitly define the conjugate state $\bra{\widehat{\Psi_{\text{dS,W}}^{\Delta_{\pm}}}(x)}$ of the bulk local state in the dual CFT$_2$.
Here, utilizing the antipodal map discussed in subsection \ref{3.4}, we explore the elegant relationship between $\bra{\widehat{\Psi_{\text{dS,W}}^{\Delta_{\pm}}}(x)}$ and $\bra{\Psi_{\text{dS,W}}^{\Delta_{\pm}}(x^{\prime})}$.
In subsection \ref{3.6}, we take a CPT-invariant linear combination of bulk local states to obtain physical states that yield the correct two-point function, and subsequently derive the two-point function in the Euclidean vacuum. 
Finally, we discuss $\alpha$-vacua in subsection \ref{3.7} and the information metric constructed from the two-point function in subsection \ref{3.8}.

\subsection{Global $\text{SL}(2,\mathbf{C})$ Virasoro generators}\label{3.1}
Let us start with the embedding in Minkowski space
\be
ds^2=-dX_{-1}^2+dX_0^2+dX_1^2+dX_2^2.
\ee
The dS$_3$ space is defined by the hypersurface 
\be
-X_{-1}^2+X_0^2+X_1^2+X_2^2=1,
\ee
where we set $R_{\rm dS}=1$.
We employ the following coordinates
\begin{align}
\begin{split}
    X_{-1}&=\cos\theta\sinh t,\quad X_0=\cos\theta\cosh t,\\
    X_1&=\sin\theta\cos\phi,\quad X_2=\sin\theta\sin\phi,  
\end{split}\label{xxxq}
\end{align}
which gives the static patches of dS$_3$
\be
    ds^2=-\cos^2{\theta} dt^2 + d\theta^2 + \sin^2{\theta} d\phi^2, \label{statds}
\ee
where $0<\theta<\pi$, $-\infty<t<\infty$ and $0<\phi<2\pi$. 
We set the dS radius $R_{\rm dS}=1$ in this paper, unless stated otherwise.
Note that this metric can be obtained by replacing the coordinates from AdS$_3$ as follows:
\be
\tau=it,\quad \rho=i\theta,\quad \phi=\phi,\quad R_{\rm AdS}^2=-R_{\rm dS}^2.\label{AdS}
\ee
In the dS/CFT correspondence \cite{Strominger:2001pn}, the dual CFT is assumed to be on the sphere at the future/past infinity $t=\pm\infty$. However, we proceed without assuming where the dual CFT lives and just assume the gravitational theory on our three-dimensional de Sitter space has a dual description in terms of a two-dimensional CFT.

The isometries of dS$_3$ form an $\mathfrak{sl}(2,\mathbf{C})$ algebra, generated by the global Virasoro generators $(L_0, L_{\pm 1})$ and $(\tilde{L}_0, \tilde{L}_{\pm 1})$ in some two-dimensional CFT. These can be explicitly written down as
\begin{align}
\begin{split}
    L_0&=\frac{1}{2}(\de_t+i\de_\phi),\quad \tilde{L}_0=\frac{1}{2}(\de_t-i\de_\phi),\\
    L_{\pm1}&=\frac{i}{2}e^{\pm (-t+i\phi)}\left[\frac{\sin\theta}{\cos\theta}\de_t-i\frac{\cos\theta}{\sin\theta}\de_\phi\mp i\de_\rho\right],\\
    \tilde{L}_{\pm1}&=\frac{i}{2}e^{\pm (-t-i\phi)}\left[\frac{\sin\theta}{\cos\theta}\de_t+i\frac{\cos\theta}{\sin\theta}\de_\phi\mp i\de_\rho\right],
\end{split}\label{visds}
\end{align}
obeying
\be
[L_m,L_n]=(m-n)L_{m+n},\quad [\tilde{L}_m,\tilde{L}_n]=(m-n)\tilde{L}_{m+n}.
\ee
From the expressions of (\ref{visds}), via a similar analysis done in section \ref{2.2} for AdS$_3$, we find that their Hermitian conjugates take the following unusual form:
\begin{align}
\begin{split}
    (L_0)^\dagger=-\tilde{L}_{0},\quad (L_{\pm1})^\dagger=\tilde{L}_{\pm1},\\
     (\tilde{L}_0)^\dagger=-L_{0},\quad (\tilde{L}_{\pm1})^\dagger=L_{\pm1}.
\end{split}\label{conjgds}
\end{align}
This is in sharp contrast with the standard conjugation (\ref{Virasoro}) obtained for AdS$_3$/CFT$_2$. As we will later see, the unusual conjugation in dS/CFT plays a crucial role. 

The geodesic length $D_{\text{dS}}(x,x')$ between two points $x=(t,\theta,\phi)$ and $x'=(t',\theta',\phi')$ can be calculated by
\ba
\cos D_{\rm dS}(x,x')=\cosh(t-t')\cos\theta\cos\theta'+\cos(\phi-\phi')\sin\theta\sin\theta'.
\ea
Note that $D_{\text{dS}}$ is real and positive for spacelike geodesics, while it becomes imaginary for timelike geodesics. For example, if we set $t=\phi=0$, it is simply given by $D_{\rm dS}=|\theta-\theta'|$. It is also useful to note that for the antipodal mapping
\ba
(X_{-1},X_0,X_1,X_2)\to (-X_{-1},-X_0,-X_1,-X_2), \label{antipds}
\ea
equivalent to $x=(t,\theta,\phi)\to x_A=(t,\pi-\theta,\phi+\pi)$ in (\ref{xxxq}), we can also show 
\ba
\cos D_{\rm dS}(x,x')=-\cos D_{\rm dS}(x,x'_A),  \label{antipdi}
\ea
which implies 
\ba
D_{\rm dS}(x,x'_A)=\pi-D_{\rm dS}(x,x'). \label{twogea}
\ea

\subsection{Bulk local states in the static coordinates of dS$_3$}\label{3.2}

Let us now consider a local excitation in the static patch of dS$_3$. We repeat the symmetry approach taken in the analysis of AdS$_3$, reviewed in the previous section. 
It is useful to introduce the $\text{SU}(2)$ generators:
\begin{align}
    \begin{split}
 J_1&=\frac{1}{2}\left[(L_1-\ti{L}_{-1})+(\ti{L}_1-L_{-1})\right],\\
 J_2&=\frac{i}{2}\left[(L_1-\ti{L}_{-1})-(\ti{L}_1-L_{-1})\right],\\
 J_3&=L_0-\ti{L}_0.  
    \end{split}\label{jsut}
\end{align}
They satisfy the $\mathfrak{su}(2)$ algebra:
\ba
[J_i,J_j]=i\ep_{ijk}J_k,
\ea
where the generators are all taken to be Hermitian $J^\dagger_i=J_i$.

\begin{figure}[hhh]
    		\centering
    \includegraphics[width=5cm]{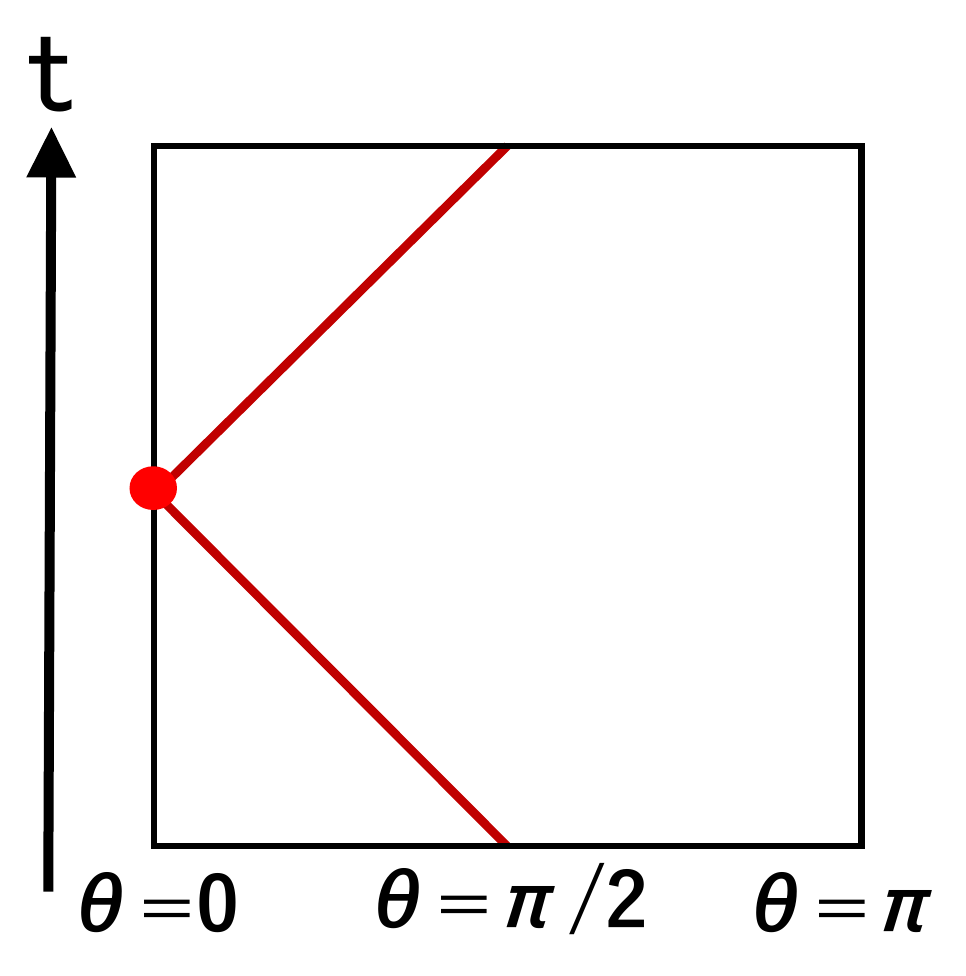}
    		\caption{The light cone $\sin\theta=|\tanh t|$ in dS$_3$, which is invariant under the actions of 
            $L_0-\tilde{L}_0$, $L_1+\tilde{L}_{-1}$, and $L_{-1}+\tilde{L}_1$.} 
    		\label{fig:dSlocal}
    \end{figure}

We first excite the point $(t,\theta,\phi)=(0,0,0)$ in (\ref{statds}) and denote its dual CFT state as $|\Psi_\Delta\lb$, where $\Delta$ labels the dimension of the primary state in the dual 2D CFT. Remembering the expressions (\ref{visds}), we see that the point $\theta=t=0$ is invariant under the actions of $L_0-\ti{L}_0$ and $L_{\pm 1}+\ti{L}_{\mp 1}$.
Therefore we require 
\ba
\begin{split}
 (L_0-\ti{L}_0)|\Psi_\Delta\lb&=0,\\
 (L_1+\ti{L}_{-1})|\Psi_\Delta\lb&=0,\\
 (L_{-1}+\ti{L}_{1})|\Psi_\Delta\lb&=0.  \end{split}\label{constrgh}
\ea
We can solve these constraints (\ref{constrgh}) by setting 
\ba
|\Psi_\Delta\lb=e^{i\frac{\pi}{2}(L_0+\ti{L}_0-\Delta)}|I_\Delta\lb,
\ea
in terms of the global Ishibashi state (\ref{Ishibashia}), as in the AdS$_3$ case (\ref{IshibashiAdS}). Using this state we would like to construct the bulk local state in the CFT dual to dS$_3$. We expect that the bulk local state is created by acting the bulk scalar field at $t=\theta=0$ and its light cone is invariant under the actions of $L_0-\tilde{L}_0$, $L_1+\tilde{L}_{-1}$, and $L_{-1}+\tilde{L}_1$, as sketched in Fig.\ref{fig:dSlocal}.

Thus we can explicitly write down the bulk local state as follows
\ba
|\Psi_\Delta\lb=\sum_{k=0}^\infty c_k(\Delta)L^k_{-1}\ti{L}^k_{-1}|\Delta\lb,
\label{bstate}
\ea
where $|\Delta\lb$ is the primary state with dimension $\Delta$ (\ref{primarycond}), and $c_k$ is given by 
\ba
c_k(\Delta)=\frac{e^{i\pi k}}{\prod_{j=1}^{k}(j^2+(\Delta-1)j)}.
\ea

In order to generalize $|\Psi_\Delta\lb$ to a generic local state which describes local excitation at an arbitrary point $(t,\theta,\phi)$, let us note that the $t=0$ slice is preserved by the $\text{SU}(2)$ generators introduced in (\ref{jsut}):
\ba
\begin{split}
J_1&=i\left(\frac{\sin\phi\cos\theta}{\sin\theta}\de_\phi-\cos\phi\de_\theta\right),\\
J_2&=i\left(\frac{\cos\phi\cos\theta}{\sin\theta}\de_\phi+\sin\phi\de_\theta\right),\\
J_3&=i\de_\phi.
\end{split}
\ea
The unitary transformation which maps the point $(t,\theta,\phi)=(0,0,0)$ to 
$(0,\theta_0,\phi_0)$ is given by $e^{-i\phi_0J_3}e^{i\theta_0J_1}$. Finally, by performing the time translation, we obtain the bulk local state 
\ba
 \ket{\Psi^{\Delta}_{\text{dS,W}}(t,\theta,\phi)}=e^{(L_0+\ti{L}_0)t}e^{-i\phi J_3}e^{i\theta J_1}|\Psi_\Delta\lb,  \label{blsdsta}
\ea
where the subscript $W$ denotes the Wick rotation.
The equation of motion for a scalar with mass $m$ can be written as 
\ba
\left[\frac{1}{2}(L_{-1}L_1+L_1L_{-1})-L_0^2+\frac{1}{2}(\ti{L}_{-1}\ti{L}_1+\ti{L}_1\ti{L}_{-1})-\ti{L}_0^2-\frac{m^2}{2}\right]\ket{\Psi^{\Delta}_{\text{dS,W}}(x)}=0,\no
\ea
where $x$ denotes the coordinates $(t,\theta,\phi)$. This fixes the conformal dimension to be 
\ba
\Delta_\pm=1\pm \s{1-m^2}\equiv 1\pm i\mu.  \label{dimret}
\ea
We therefore have two different bulk local states  
\ba
\ket{\Psi^{\Delta_\pm}_{\text{dS,W}}(x)}
=e^{(L_0+\ti{L}_0)t}e^{-i\phi J_3}e^{i\theta J_1}\sum_{k=0}^\infty c_k(\Delta_{\pm})L^k_{-1}\ti{L}^k_{-1}|\Delta_{\pm}\lb,  \label{localds}
\ea

However, the bulk local states in dS$_3$ constructed above are obtained simply by Wick rotating those in AdS$_3$ and do not correspond to physical states because we cannot obtain the correct two-point function using $\ket{\Psi^{\Delta_\pm}_{\text{dS,W}}(x)}$. 
We will construct physical states later in this section. Before turning to that, let us evaluate the inner product of the Wick-rotated bulk local states.

\subsection{Two-point functions via Wick rotation from AdS$_3$}\label{3.3}
Now we would like to turn to the inner product of the locally excited state.
We consider the Wick rotation of the two-point function in AdS$_3$ (\ref{AdStwop}). This leads to 
\begin{align}
    \braket{\Psi_{\text{dS,W}}^{\Delta_{\pm}}(x^{\prime})|\Psi_{\text{dS,W}}^{\Delta_{\pm}}(x)}=\frac{e^{\pm\mu D_{\text{dS}}(x,x')}}{2i\sin D_{\text{dS}}},\label{innerds}
\end{align}
where the conformal dimension is given by (\ref{dimret})\footnote{Similar form is found in \cite{Sleight:2021plv}.}.
For this, we should note the coordinate transformation (\ref{AdS}) and the relation
\ba
D_{\text{AdS}}(x,x')=iD_{\text{dS}}(x,x'),\ \ \label{DDs}
\ea
so that we have $D_{\text{dS}}(x,x')=\theta$ when $x=(0,\theta,0)$ and $x'=(0,0,0)$.
Here, the bra state $\bra{\Psi_{\text{dS,W}}^{\Delta_{\mp}}(x^{\prime})}$ is defined via an analytic continuation from the AdS case as
\ba
\bra{\Psi_{\text{dS,W}}^{\Delta_{\pm}}(x^{\prime})}=\la \Delta_{\pm}|
\left(\sum_{k=0}^\infty c_k(\Delta_{\pm}) L^k_1\ti{L}^k_1\right) e^{-i\theta J_1}e^{i\phi J_3}e^{-(L_0+\ti{L}_0)t},\label{bradslo}
\ea
where we should note that the coefficient $c_k(\Delta_\pm)$ in (\ref{bstate}) becomes complex-valued for (\ref{dimret}) and satisfies $c^*_k(\Delta_{\pm})=c_k(\Delta_\mp)$. This bra state is introduced so that the inner product leads to the Wick rotation of the AdS$_3$ result:
\ba
\la \Psi(t,\theta)|\Psi(0,0)\lb=\frac{1}{(\cos\theta)^{\Delta}}
\sum_{k=0}^\infty e^{-(2k+\Delta)t}P^{(0,\Delta-1)}_k\left(\frac{2}{\cos^2\theta}-1\right).  \label{dSsum}
\ea

In the calculation above, we introduced the bra state via a naive Wick rotation. It is not immediately clear if it is related to the bra state obtained by applying the unusual conjugation in dS$_3$ (\ref{conjgds}) on the corresponding ket state. Indeed as will show shortly, the conjugation does not map $\ket{\Psi^{\Delta_\pm}_{\text{dS,W}}(x)}$ to $\bra{\Psi_{\text{dS,W}}^{\Delta_{\pm}}(x^{\prime})}$, but instead to $\bra{\Psi_{\text{dS,W}}^{\Delta_{\mp}}(x^{\prime}_A)}$ up to a constant factor.

\subsection{Antipodal map}\label{3.4}
It is useful to examine the antipodal map. We can conveniently describe the antipodal map $x\to x_A$ (\ref{antipds}) by the action $(t,\theta,\phi)\to (t,\theta+\pi,\phi)$. The bulk local state at $t=\theta=0$ (\ref{bstate}) transforms as follows
\ba
\ket{\Psi_{\Delta_\pm}}\to e^{i\pi J_1}\ket{\Psi_{\Delta_\pm}}.
\ea
Since this map $e^{i\pi J_1}$ has the following action on the Virasoro generators
(see appendix \ref{ap:formula}),
\ba
\begin{split}
 e^{i\pi J_1}L_0 e^{-i\pi J_1}&=-L_0,\\
 e^{i\pi J_1}L_{\pm 1} e^{-i\pi J_1}&=-L_{\mp 1},\\
 e^{i\pi J_3}L_{\pm 1} e^{-i\pi J_3}&=-L_{\pm 1},
\end{split}\label{Jpi}
\ea
it is easy to see that the state after being acted on by the antipodal map satisfies the same condition (\ref{constrgh}). This shows that the antipodal state is proportional to the original state:
\ba
\ket{\Psi^{\Delta_\pm}_{\text{dS,W}}(t,\theta,\phi)}\propto\ket{\Psi^{\Delta_\pm}_{\text{dS,W}}(t,\theta+\pi,\phi)}.
\ea
At the point $t=\theta=\phi=0$, this is written as
\ba
e^{i\pi J_1}\ket{\Psi_{\Delta_\pm}}=\lambda_{\pm}\ket{\Psi_{\Delta_\pm}},
\label{joneds}
\ea
where $\lambda_\pm$ are the proportionality coefficients.
Since (\ref{innerds}) leads to
\ba
 \braket{\Psi_{\text{dS,W}}^{\Delta_{\pm}}(0,\pi+\theta,0)|\Psi_{\text{dS,W}}^{\Delta_{\pm}}(0,0,0)}=-e^{\pm\pi\mu}\braket{\Psi_{\text{dS,W}}^{\Delta_{\pm}}(0,\theta,0)|\Psi_{\text{dS,W}}^{\Delta_{\pm}}(0,0,0)},
\ea
we find
\ba
\lambda_{\pm}=-e^{\pm \pi\mu}.
\ea

Using (\ref{constrgh}) we can further rewrite (\ref{joneds})  as
\ba
e^{i\pi (L_1-L_{-1})}\ket{\Psi_{\Delta_\pm}}=\lambda_{\pm}\ket{\Psi_{\Delta_\pm}}.
\ea
Since $e^{i\pi (L_1-L_{-1})}$ commutes with $L_{0}$ and $L_{\pm 1}$, this identity is equivalent to that for the primary state:
\ba
e^{i\pi (L_1-L_{-1})}|\Delta_\pm\lb=\lambda_{\pm}|\Delta_\pm\lb.
\ea

Thus we obtain
\ba
\la \Delta_\pm|e^{-i\pi(L_1-L_{-1})}|\Delta_\pm\lb=\la \Delta_\pm|e^{-i\pi(\ti{L}_1-\ti{L}_{-1})}|\Delta_\pm\lb=-e^{\pm \pi\mu}.  \label{innerjone}
\ea
The notation is summarized in the 
section $\ref{Notation}$.

\subsection{de Sitter conjugation}\label{3.5}

Now to better understand the definition of bra states (\ref{bradslo}), we would like to examine how states transform under the conjugation (\ref{conjgds}) unique to the CFT dual to dS$_3$. First and foremost, conjugations of primary states $\ket{\Delta_\pm}$, which satisfy (\ref{primarycond}), obey
\ba
\begin{split}
    &|\Delta_\pm\lb^\dagger L_0^\dagger =\frac{\Delta_\mp}{2}|\Delta_\pm\lb^\dagger,
    \qquad|\Delta_\pm\lb^\dagger L_1^\dagger=0,\\
   &L_0^\dagger\bra{\Delta_\pm}^\dagger=
   \bra{\Delta_\pm}^\dagger\frac{\Delta_\mp}{2},\qquad
   L_{-1}^\dagger\bra{\Delta_\pm}^\dagger=0.
\end{split}
\ea
The same identities hold for the anti-chiral sector.
Thus, if we write 
\ba
|\Delta_\pm\lb^\dagger\equiv\la \widehat{\Delta_{\pm}}|,  \quad \bra{\Delta_{\pm}}^\dagger=\ket{\widehat{\Delta_{\pm}}},\label{dualdsa}
\ea
these satisfy the following exotic properties 
\ba\label{wwees}
\begin{split}
    &\la \widehat{\Delta_{\pm}}| L_0=-\frac{\Delta_\mp}{2}\la \widehat{\Delta_{\pm}}|,\quad \mbox{and}\quad
    \la \widehat{\Delta_{\pm}}| L_1=0,\\
    &L_0\ket{\widehat{\Delta_\pm}}=-\frac{\Delta_\mp}{2}\ket{\widehat{\Delta_\pm}},\quad\text{and}\quad
    L_{-1}\ket{\widehat{\Delta_\pm}}=0.
\end{split}
\ea
\eqref{wwees} significantly differs from the AdS case \cite{Miyaji:2015fia}, but can be understood by treating this CFT as a non-Hermitian system.
By regarding this CFT as a non-Hermitian system, the following normalization can be adopted\footnote{For more details, refer to appendix \ref{appendix_nonhemitian}}:
\begin{align}\label{normalization_dS_3}
    \begin{split}
        &\braket{\Delta_i|\Delta_j}=\braket{\widehat{\Delta_i}| \widehat{\Delta_j}}=\delta_{ij},\\
        &\braket{\Delta_i|\widehat{\Delta_j}}
        =\braket{\widehat{\Delta_i}|\Delta_j}=0.
    \end{split}
\end{align}

Then the conjugations of bulk local states $\ket{\Psi^{\Delta_\pm}_{\text{dS,W}}(x)}$ read 
\ba
\ket{\Psi^{\Delta_\pm}_{\text{dS,W}}(x)}^{\dagger}
\equiv &&\bra{\widehat{\Psi_{\text{dS,W}}^{\Delta_{\pm}}}(x)}\nonumber\\
=&&\la \widehat{\Delta_\pm}|\sum_{k}c_k(\Delta_\mp) L^k_{-1}\ti{L}^k_{-1} e^{-i\theta J_1}e^{i\phi J_3}e^{-(L_0+\ti{L}_0)t},
\ea
where in the first line we introduced the expression of the conjugated state $\bra{\widehat{\Psi_{\text{dS,W}}^{\Delta_{\mp}}}(x)}$.
Next we insert $1=e^{i\pi J_1}\cdot e^{-i\pi J_1}$ in the above expression and rewrite as follows:
\ba
\bra{\widehat{\Psi_{\text{dS,W}}^{\Delta_{\pm}}}(x)}&=&\la \widehat{\Delta_\pm}|\sum_{k}c_k(\Delta_\pm)L^k_{-1}\ti{L}^k_{-1} e^{i\pi J_1}\cdot e^{-i\pi J_1}e^{-i\theta J_1}e^{i\phi J_3}e^{-(L_0+\ti{L}_0)t}\no
&=&\la \widehat{\Delta_\pm}|e^{i\pi J_1}\sum_{k}c_k(\Delta_\mp) L^k_{1}\ti{L}^k_{1} \cdot e^{-i(\theta+\pi) J_1}e^{i\phi J_3}e^{-(L_0+\ti{L}_0)t}.
\ea

Now, to proceed further, note the following properties can be found from (\ref{wwees}):
\begin{align}
 &\la \widehat{\Delta_\pm}|e^{i\pi J_1}L_0=\frac{\Delta_\mp}{2}\cdot\la \widehat{\Delta_\pm}|e^{i\pi J_1},\nonumber\\
 &\la\widehat{\Delta_\pm}|e^{i\pi J_1}L_{-1}=0,
\end{align}
which shows that $\la \widehat{\Delta_\pm}|e^{i\pi J_1}$ obey the condition of standard primary states.
Since we do not expect primary states dual to the scalar field other than $|\Delta_\mp\lb$, we expect that $\la \widehat{\Delta_\pm}|e^{i\pi J_1}$ is proportional to $\la \Delta_\mp|$:
\ba
&& \la \widehat{\Delta_\pm}|e^{i\pi J_1}=\nu_\mp\la \Delta_\mp|.
\ea
By taking their conjugations with (\ref{dualdsa}), we obtain
\ba
&& e^{-i\pi J_1}|\Delta_\pm\lb=\nu^*_\mp |\widehat{\Delta_\mp}\lb.
\ea
Then, we find
\ba
\nu_\pm\cdot \la \Delta_\pm|e^{-2i\pi J_1}|\Delta_\pm\lb=\nu^*_\mp. \label{nuconstraint}
\ea
Our previous result (\ref{innerjone}) shows $\la \Delta_\pm|e^{-2i\pi J_1}|\Delta_\pm\lb=e^{\pm 2\pi \mu}$, by taking into account both contributions from the chiral and anti-chiral sectors. As the simplest solution to (\ref{nuconstraint}), we choose
\ba
\nu_\pm=e^{\mp \pi\mu}.  \label{solnu}
\ea

This way we found that the conjugated states $|\Psi_{\pm}(x)\lb^{\dagger}$ are given by the antipodal transformation of the expected bra states (\ref{bradslo}) 
\begin{align}\label{condspm}
\bra{\widehat{\Psi_{\text{dS,W}}^{\Delta_{\pm}}}(x)} &=\nu_\mp\la \Delta_\mp|\sum_{k}c_k(\Delta_\mp) L^k_{1}\ti{L}^k_{1} \cdot e^{-i(\theta+\pi) J_1}e^{i\phi J_3}e^{-(L_0+\ti{L}_0)t} \nonumber\\
&=\nu_\mp\bra{\Psi_{\text{dS,W}}^{\Delta_{\mp}}(t,\theta+\pi,\phi)}.
\end{align}
We should note that the inner product between $|\Psi_{\pm}(x)\lb^{\dagger}$ and $|\Psi_{\pm}(x)\lb$ is vanishing, while the one between $|\Psi_{\mp}(x)\lb^{\dagger}$ and $|\Psi_{\pm}(x)\lb$ is non-vanishing. This is due to the normalization \eqref{normalization_dS_3}, which reflects the non-Hermitian nature of the CFT dual to dS$_3$.

Finally the inner product of bulk local states and their conjugations is evaluated as
\ba\label{twopointfunction_via_conjugation}
\braket{\widehat{\Psi_{\text{dS,W}}^{\Delta_{\mp}}}(x)|\Psi^{\Delta_\pm}_{\text{dS,W}}(x')}
=\frac{e^{\mp\mu\left(\pi- D_{dS}(x_A,x')\right)}}{2i\sin D_{dS}(x_A,x')}
=\frac{e^{\mp\mu D_{dS}(x,x')}}{2i\sin D_{dS}(x,x')},  \label{twopointpm}
\ea
while we have $\braket{\widehat{\Psi_{\text{dS,W}}^{\Delta_{\pm}}}(x)|\Psi^{\Delta_\pm}_{\text{dS,W}}(x')}=0$.

Note that if we focus only a single primary with dimension $\Delta_+$ (or $\Delta_{-}$), then we cannot obtain any meaningful two-point functions as the inner product between the state and its conjugate is vanishing. To obtain a non-trivial result, we need to mix two primaries 
$\Delta_+$ and $\Delta_-$. In the next subsection, we will be able to reproduce the expected two-point function in dS$_3$ by combining them and using (\ref{twopointpm}). The correct two-point function (Wightman function) in the Euclidean (Bunch-Davies) vacuum \cite{Bunch:1978yq,Hartle:1983ai} takes the following form  (see e.g. \cite{Strominger:2001pn}):
\begin{align}
    G_{\rm E}(x,x')&=\frac{\Gamma[\Delta_+]\Gamma[\Delta_-]}{(4\pi)^{\frac{3}{2}}\Gamma\left[\frac{3}{2}\right]}{}_2F_1\left(\Delta_+,\Delta_-,\frac{3}{2};\cos^2\frac{D_{\rm dS}}{2}\right)\nn\\
    &=\frac{\sinh\mu(\pi-D_{\rm dS}(x,x'))}{4\pi \sinh \mu\pi \sin D_{\rm dS}(x,x')}.\label{twopdse}
\end{align}
This two-point function becomes divergent only when the two points coincide $x=x'$, while the one obtained from the naive Wick rotation from AdS has another divergence at the antipodal point $x'=x_A$.

\subsection{Two-point functions in the dS Euclidean vacuum}\label{3.6}

We argue that in order to have physical excitation in dS$_3$, we should take a special linear combination of bulk local states corresponding to dimensions $\Delta_+$ and $\Delta_-$. 

First we note the relation $\phi_{\text{AdS}}=\pm\s{i}\phi_{\text{dS}}$ for the scalar field in dS$_3$. This is due to the kinetic term of the bulk scalar in dS$_3$ looking like 
\be
S=R_{\rm AdS}\int \de^\mu\phi_{\rm AdS}\de_\mu\phi_{\rm AdS}=R_{dS}\int \de^\mu\phi_{\rm dS}\de_\mu\phi_{\rm dS},
\ee
as well as the relationship between the dS and AdS radii given by $R^2_{\rm dS}=-R^2_{\rm AdS}$. Having this in mind, we argue that the physical local excitation by the scalar field in the Euclidean vacuum is dual to the following state in the CFT:
\ba
|\Psi_{\text{E}}(x)\lb=\frac{1}{2\s{\pi\sinh\pi\mu}}\left[\frac{1}{\s{i}}\ket{\Psi^{\Delta_+}_{\text{dS,W}}(x)}
+\s{i}\ket{\Psi^{\Delta_{-}}_{\text{dS,W}}(x_A)}\right],  \label{psiEa}
\ea
where we chose the overall constant so that it agrees with the normalization of the two-point function (\ref{twopdse}). See also \cite{Xiao:2014uea,Sleight:2020obc,Sleight:2021plv} for closely related linear combinations in terms of bulk fields. 
Notice that this state satisfies the equation of motion of a scalar field 
\ba
\left[\frac{1}{2}(L_{-1}L_1+L_1L_{-1})-L_0^2+\frac{1}{2}(\ti{L}_{-1}\ti{L}_1+\ti{L}_1\ti{L}_{-1})-\ti{L}_0^2-\frac{m^2}{2}\right]\ket{\Psi_E(x)}=0.
\ea
We observe that the state $|\Psi_{\text{E}}(x)\lb$ is invariant under CPT transformation, where the coordinates are mapped to the antipodal point $x\to x_A$.
Since this is an antilinear map, the conformal dimensions $\Delta_{\pm}=1\pm i\mu$ are transformed into their complex conjugates $\Delta_{\mp}=1\mp i\mu$. Therefore the CPT transformation acts on the bulk local states as 
$\ket{\Psi^{\Delta_\pm}_{\text{dS,W}}(x)}\to \ket{\Psi^{\Delta_{\mp}}_{\text{dS,W}}(x_A)}$. We should also mention that gauging CPT symmetry in de Sitter space appears to play an important role in recent works \cite{Harlow:2023hjb,Susskind:2023rxm,Goodhew:2024eup}.

By taking its conjugation using (\ref{condspm}), we obtain
\ba
\la \Psi_{\rm E}(x)|=\frac{1}{2\s{\pi\sinh\pi\mu}}\left[\s{i}\bra{\widehat{\Psi_{\text{dS,W}}^{\Delta_{+}}}(x)}+\frac{1}{\s{i}}
\bra{\widehat{\Psi_{\text{dS,W}}^{\Delta_{-}}}(x_A)}\right].  \label{psiEb}
\ea

Finally the inner product of  (\ref{psiEa}) and  (\ref{psiEb}) matches with the expected physical two-point function (\ref{twopdse}):
\ba
 \la \Psi_{\rm E}(x)|\Psi_{\rm E}(x')\lb&=&\frac{i}{4\pi\sinh\pi\mu}\left[\braket{\widehat{\Psi_{\text{dS,W}}^{\Delta_{+}}}(x)|\Psi^{\Delta_{-}}_{\text{dS,W}}(x'_A)}-\braket{\widehat{\Psi_{\text{dS,W}}^{\Delta_{-}}}(x_A)|\Psi^{\Delta_{+}}_{\text{dS,W}}(x')}\right] \no
&=& \frac{i}{4\pi\sinh\pi\mu}\left[\frac{e^{\mu\left(\pi-D_{\rm dS}(x,x')\right)}}{2i\sin D_{\rm dS}(x,x')}-\frac{e^{-\mu\left(\pi-D_{\rm dS}(x,x')\right)}}{2i\sin D_{\rm dS}(x,x')}\right]\no
&=&\frac{\sinh\mu\left(\pi-D_{\rm dS}(x,x')\right)}{4\pi \sinh \mu\pi \sin D_{\rm dS}(x,x')}.\label{GEds}
\ea
For a fixed value of $x$, this two-point function becomes divergent only when $x'=x$ and this confirms that the state $|\Psi_{\rm E}(x')\lb$ is dual to the bulk local state where the excitation is localized at the point $x'$. Moreover it agrees with the expected two-point function in the bulk dS$_3$.

We shall make a comment on the case where the conformal dimension takes on real values. In this case, as expected from analytic continuation $i\mu\rightarrow \gamma$, the Wightman function in the Euclidean vacuum reads
\be
G_{\rm E}(x',x)=\frac{\sin \gamma(\pi-D_{\rm dS})}{4\pi\sin \gamma\pi \sin D_{\rm dS}},\label{real}
\ee
where $\gamma=\sqrt{m^2-1}\in \mathbf{R}$. In the usual dS/CFT setup \cite{Strominger:2001pn}, for real conformal dimensions there are fast and slowly falling modes, as in the AdS/CFT case, and we only need normalizable modes to obtain the Green's function. However, (\ref{real}) signals that we need both modes since there is a sine factor in the numerator. Thus, we expect that for real conformal dimensions we can use the same argument with the replacement $i\mu=\gamma$.

\subsection{$\alpha$-vacua}\label{3.7}

It is known that the vacuum state, which is invariant under the symmetries of de Sitter space, is not unique. The symmetry-invariant vacua are parameterized by real-valued $\ap$, and are therefore called the $\alpha$-vacua. Here we would like to present the bulk local state that reproduces the Green's function of $\alpha$-vacua given in \cite{Bousso:2001mw}.

The bulk local state in the $\alpha$-vacua is written by
\be
\ket{\Psi_\alpha(x)}=\frac{1}{\sqrt{1-e^{2\alpha}}}(\ket{\Psi_E(x)}+e^\alpha\ket{\Psi_E(x_A)}).
\ee
It is clear that this satisfies the symmetric constraints (\ref{constrgh}) and thus is invariant under the symmetries of de Sitter space. Then, the two-point function reads
\begin{align}
    &\braket{\Psi_\alpha(x)|\Psi_\alpha(x')}\nonumber\\
    =&\frac{1}{1-e^{2\alpha}}\left(G_E(x,x')+e^\alpha G_E(x,x_A')+e^{\alpha}G_E(x_A,x')+e^{2\alpha}G_E(x_A,x_A')\right),
\end{align}
which reproduces the result in \cite{Bousso:2001mw}.

\subsection{dS$_3$ from the information metric}\label{3.8}

Finally we would like to examine how the holographic spacetime looks from the viewpoint of the bulk local state $|\Psi_{\rm E}(x)\lb$ as a probe. We will eventually find that the metric it probes is given by the dS$_3$ metric, which is a dS generalization of the result for AdS$_3$ in \cite{Miyaji:2015fia}. 
Even though this result naturally follows from the conformal symmetry of our CFT construction, we will briefly present this argument below as another supporting evidence of the emergence of dS$_3$ spacetime from the 2D CFT. 

Since this state $|\Psi_{\rm E}(x)\lb$ 
is singular in that it is localized at a point $x$ and has infinite energy and a divergent inner product (indeed the two-point function (\ref{GEds}) becomes divergent when $x=x'$), we regularize the divergence by smearing the bulk local excitation around the point $x$ over a region with a small size $\delta$. By introducing such a short scale cutoff $\delta$ that preserves the conformal symmetry, we expect that the two-point function to have the following behavior in the limit $x\to x'$\footnote{Though we do not explicitly construct the regularized state, it is well approximated by the state $e^{-\delta (L_0+\ti{L}_0)}|\Psi_{\rm E}(x)\lb$. 
Indeed this was used in \cite{Miyaji:2015fia}
to reproduce the AdS$_3$ metric. The same argument can be applicable to the current dS$_3$ case. We expect that the localization of this state is smeared around the region with size $\delta$.}:
\ba
 \la \Psi_{\rm E}(x)|\Psi_{\rm E}(x')\lb \simeq \frac{\delta}{\s{D_{\rm dS}(x,x')^2+\delta^2}},
\ea
Here we normalized the bulk local state so that we have $ \la \Psi_{\rm E}(x)|\Psi_{\rm E}(x)\lb=1$. 

The information metric $G_{ij}$ (or Bures distance) is defined by 
\ba
G_{ij}dx^i dx^j=1-|\la \Psi_E(x)|\Psi_E(x+dx)\lb|.
\ea
Since we have
\ba
D^2_{dS}(x,x+dx)\simeq -\cos^2\theta dt^2+d\theta^2+\sin^2\theta d\phi^2,
\ea
we obtain the information metric
\ba
ds_{\rm inf}^2=\frac{1}{2\delta^2}\left(-\cos^2\theta dt^2+d\theta^2+\sin^2\theta d\phi^2\right). \label{dsmetrgw}
\ea
In this way we find that the information metric is proportional to that of dS$_3$. Notice that we successfully obtain a timelike metric for the $t$-coordinate. The crucial reason why we find the timelike coordinate from bulk local states is because of the unusual conjugation operation 
(\ref{conjgds}) that flips the sign of energy $L_0+\ti{L_0}$.

Quantum estimation theory tells us that the inverse of information metric $G$ provides the lower bound of quantum uncertainty of the coordinates $x$ such that $\la \delta x_i\delta x_j \lb\geq (G^{-1})_{ij}$, when we try to estimate their values from quantum measurements. Refer to \cite{Miyaji:2015yva} for more details of the application of quantum estimation theory to AdS/CFT. This means that the UV cutoff scale $\delta$ describes the minimal distance where we can trust the dS$_3$ metric (\ref{dsmetrgw}). For scales smaller than $\delta$, the uncertainty washes out the classical spacetime. If we assume that the estimation done for AdS/CFT in \cite{Miyaji:2015fia} can also be applied to our dS$_3/$CFT$_2$, it is natural to identify $\delta$ as $\frac{1}{\delta}\sim \frac{R_{\rm dS}}{G_{\rm N}}$, where $R_{\rm dS}$ is the dS$_3$ radius and $G_{\rm N}$ is the three-dimensional Newton constant. This might reproduce the expected dS$_3$ metric up to an $\mathcal{O}(1)$ factor.

\section{Global coordinates in dS$_3$}\label{Sec:dSglobal}
To understand how this holographic duality looks in the full de Sitter spacetime, it would be useful to repeat the analysis using the global coordinates $(T,\psi,\phi)$ of dS$_3$. The metric looks like
\begin{align}
    ds^2=-dT^2+\cosh^2{T}\left(d\psi+\sin^2{\psi}d\phi^2\right)
\end{align}
In global dS$_3$, the geodesic distance $D_{\rm dS}$ reads
\ba
 \cos D_{\rm dS}(x,x') &=&\eta_{ab}X^aX'^b,\no
&=& -\sinh T\sinh T'+\cosh T\cosh T'\left(\cos\psi\cos\psi'+\sin\psi\sin\psi'\cos(\phi-\phi')\right).\no
\ea
The coordinate transformation between the static patch and global coordinates is given by 
\ba
\begin{split}
X_{-1}&=\sinh T=\cos\theta\sinh t,\\
X_0&=\cosh T\cos\psi=\cos\theta\cosh t,\\
X_1&=\cosh T\sin\psi\cos\phi=\sin\theta\cos\phi,\\
X_2&=\cosh T\sin \psi \sin\phi=\sin\theta\sin\phi.
\end{split}
\ea

The $\text{SL}(2,\mathbf{C})$ generators defined by (\ref{visds}) in the static patch is rewritten in terms of the global coordinates:
\ba
\begin{split}
L_0&=\frac{1}{2}\left(\cos\psi\de_T-\sin\psi\tanh T\de_\psi+i\de_\phi\right),\\
\ti{L}_0&=\frac{1}{2}\left(\cos\psi\de_T-\sin\psi\tanh T\de_\psi-i\de_\phi\right),\\
L_{\pm 1}&=\frac{i}{2}e^{\pm i\phi}\left[\sin\psi\de_T-i\left(\frac{\cos\psi\mp\tanh T}{\sin\psi}\right)\de_\phi+(\cos\psi\tanh T\mp 1)\de_\psi\right],\\
\ti{L}_{\pm 1}&=\frac{i}{2}e^{\mp i\phi}\left[\sin\psi\de_T+i\left(\frac{\cos\psi\mp\tanh T}{\sin\psi}\right)
\de_\phi+(\cos\psi\tanh T\mp 1)\de_\psi\right].
\end{split}
\ea

\subsection{Bulk local states in global coordinates}

If we require that the action preserves the point $\psi=0$ and $T=T_0$, we obtain the previous constraints (\ref{constrgh}) for the corresponding bulk local state. By introducing the $\mathfrak{su}(2)$ algebra (\ref{jsut}) again, we can show at any point $(T,\psi,\phi)$:
\ba
\begin{split}
J_1&=-i\cos\phi\de_\psi+i\sin\phi\frac{\cos\psi}{\sin\psi}\de_\phi, \\
J_3&=i\de_\phi,\\
L_0+\ti{L}_0&=\cos\psi \de_T-\sin \psi \tanh T\de_\psi.
\end{split}
\ea
Near the point $\psi=0$, if we define $z=\psi e^{i\phi}$ and $\bar{z}=\psi e^{-i\phi}$, we find 
$J_1\simeq -2i(\de_z+\de_{\bar{z}})$. Therefore $J_1$ shifts the point $\psi=0$ in the (horizontal) $\phi=0$ direction.

Finally the bulk local state in global dS$_3$ at $(T,\psi,\phi)=(T_0,\psi_0,\phi_0)$ with conformal dimension $\Delta_\pm$ (\ref{dimret}) is found to be
\ba
\ket{\Psi^{\Delta_{\pm}}_{\text{dS(global),W}}(T_0,\psi_0,\phi_0)}=e^{-i\phi_0 J_3}e^{i\psi_0J_1}
e^{(L_0+\ti{L}_0)T_0}|\Psi_{\Delta_{\pm}}\lb,  \label{gglobal}
\ea
where $|\Psi_{\Delta_{\pm}}\lb$ is identical to (\ref{bstate}). 

\subsection{Relation to bulk local states in static dS$_3$}

We argue that the bulk local states in global dS$_3$ (\ref{gglobal}) are equivalent to those in static dS$_3$ (\ref{blsdsta}):
\ba
\ket{\Psi^{\Delta_{\pm}}_{\text{dS(global),W}}(T_0,\psi_0,\phi_0)}= \ket{\Psi^{\Delta_{\pm}}_{\text{dS,W}}(t_0,\theta_0,\phi_0)},
\label{idends}
\ea
by the obvious identification 
\ba
\begin{split}
\sinh T_0&=\cos\theta_0\sinh t_0,\\
\cosh T_0\cos\psi_0&=\cosh t_0\cos\theta_0,\\
\cosh T_0\sin\psi_0&=\sin\theta_0.
\end{split}
\ea
Since $\phi$ rotation on both sides can be cancelled, it is sufficient to show
\ba
e^{i\psi_0J_1}e^{(L_0+\bar{L}_0)T_0}|\Psi_{\Delta_{\pm}}\lb
=e^{(L_0+\bar{L}_0)t_0}e^{i\theta_0J_1}|\Psi_{\Delta_{\pm}}\lb.\label{glstp}
\ea
Moreover, by using the property (\ref{constrgh}) of the state $|\Psi_{\Delta_{\pm}}\lb$, we see that (\ref{glstp}) is equivalent to
\ba
e^{i\frac{\psi_0}{2}(L_1-L_{-1})}e^{2T_0L_0}e^{i\frac{\psi_0}{2}(L_1-L_{-1})}|\Psi_{\Delta_{\pm}}\lb
=e^{t_0L_0}e^{i\theta_0(L_1-L_{-1})}e^{t_0L_0}|\Psi_{\Delta_{\pm}}\lb.
\ea
This can be easily demonstrated by replacing the generators by Pauli matrices (see appendix \ref{ap:pauli} for justification) as follows:
\be
\frac{1}{2}(L_1-L_{-1})\to \frac{1}{2}\sigma_1,  \ \ \ 
\frac{i}{2}(L_1+L_{-1})\to \frac{1}{2}\sigma_2,  \ \ \ 
 L_0\to \frac{1}{2}\sigma_3.
\ee
The identity
\ba
e^{i\frac{\psi_0}{2}(L_1-L_{-1})}e^{2T_0L_0}e^{i\frac{\psi_0}{2}(L_1-L_{-1})}=e^{t_0L_0}e^{i\theta_0(L_1-L_{-1})}e^{t_0L_0}
\label{global_static_identity}
\ea
holds, completing the proof of (\ref{idends}). 

The above result shows that bulk local states can be extended successfully to the full global dS$_3$. To construct the bulk local state which describes the Euclidean vacuum of dS$_3$, we can take the linear combination given in (\ref{psiEa}) and (\ref{psiEb}). It is clear that this reproduces the correct two-point function in global dS$_3$. Notice that in our approach, only a single CFT is enough to reproduce correlation functions in global dS$_3$, though there are two spacelike boundaries.

\section{Conclusion and discussion}\label{Sec:CD}
In this paper, we constructed CFT states dual to bulk local excitations in dS$_3$, called the bulk local states.
The isometry group of dS$_3$ is $\text{SL}(2,\mathbf{C})$, which can be identified with the global Virasoro generators $(L_0, L_{\pm 1})$ and $(\tilde{L}_0, \tilde{L}_{\pm 1})$ in a two-dimensional CFT.
Initially, we constructed bulk local states that respect the symmetry at the coordinate origin, in a manner similar to the AdS$_3$ analysis \cite{Miyaji:2015fia}. However, we soon discovered that states \eqref{blsdsta} obtained through a simple analytic continuation from AdS$_3$ were not at all physical. This is demonstrated by the failure of analytic continuation of the two-point function from AdS$_3$ \eqref{innerds} to accurately reproduce the physical two-point function in dS$_3$. The main reason for this is that the conjugation operation in dS$_3$/CFT$_2$ is fundamentally different from that in AdS$_3$/CFT$_2$ \eqref{conjgds}. By thoroughly analyzing the conjugation of states, we explained that the correct conjugation in de Sitter space partially involves antipodal operations, which arise from a nontrivial normalization factor \eqref{solnu}. Nevertheless, even the inner product yielded from dS$_3$ conjugation \eqref{twopointfunction_via_conjugation} did not return the expected two-point function.

Instead, we introduced the Euclidean vacuum state $\ket{\Psi_{\rm E}(x)}$, which is invariant under the CPT transformation, by taking a special linear combination of the bulk local states corresponding to conformal dimensions $\Delta_+$ and $\Delta_-$. Using these states, we showed that it perfectly agrees with the known two-point function in the Euclidean vacuum. We also presented an $\alpha$-vacua extension of bulk local states. Moreover, using our bulk local states, we computed the information metric for the quantum estimation of bulk coordinate values and showed that it reproduces the dS$_3$ metric. This provides an explanation for how a timelike coordinate emerges from the dual Euclidean CFT. 

We can better see the timelike nature of the $t$-coordinate in (\ref{localds}) as follows. Our argument in this paper lies on the basic assumption of dS/CFT that the dual of dS$_3$ is a two-dimensional Euclidean CFT, whose conformal symmetries match with the isometries of dS$_3$ \cite{Strominger:2001pn}. Operating under this assumption, we obtained the bulk local state (\ref{localds}), where $t$ is originally introduced as Euclidean time, as evident from the expression $e^{t(L_{0}+\ti{L}_0)}$. However, due to the unusual choice of conjugation in the present paper, $L_{0}+\ti{L}_0$ is anti-Hermitian, effectively making the parameter $t$ in the bulk local state behave like Lorentzian time. This provides a mechanism for the emergence of Lorentzian time from Euclidean CFT.
This mechanism also seems to fit nicely with the argument in \cite{Doi:2022iyj} that reduced density matrices in such CFTs become non-Hermitian and their entropy should properly be viewed as pseudoentropy \cite{Nakata:2020luh} instead of entanglement entropy. The imaginary part of pseudoentropy corresponds to the emergent time. Indeed the unconventional conjugation we introduced in this paper implies that reduced density matrices are non-Hermitian.\\

We discuss future directions of this work below.

{\bf Bulk locality}

We expect that the bulk local state we constructed should actually have a non-locality $\delta$ with respect to the values of the bulk coordinates $(t,\theta,\phi)$, depending on the degrees of freedom and the strength of the interactions in the dual CFT. In the case of AdS$_3/$CFT$_2$, we expect that, in the large central charge and strong coupling limit, the non-locality scales like $\delta\sim 1/c$. However if we consider a weakly interacting CFT, we expect a much larger non-locality $\delta\sim 1$. It would be very useful to do a similar analysis for our dS$_3/$CFT$_2$. In order for the central charge to be involved, we expect that the full Virasoro algebra will play an important role.

{\bf Bulk reconstruction}

As noted previously, there is no notion of fast and slowly falling modes in the case of complex conformal dimensions. If we apply the usual HKLL reconstruction method \cite{Hamilton:2005ju,Hamilton:2006az}, the mode sum approach is ambiguous. However, we can still reconstruct the operators using Green's theorem, though it is integrated over timelike separated regions. This breaks causality (two spacelike-separated operators do not commute) as summarized in \cite{Xiao:2014uea}. It will be intriguing to address the problem by rewriting our formula as an integration of operators. For discussions on HKLL for dS/CFT, refer to 
\cite{Chatterjee:2015pha,Chatterjee:2016ifv}.

{\bf Non-Hermitian realization}

We conducted our analysis by leveraging the correspondence between the dS dual and Euclidean CFT as a non-Hermitian system \cite{Bender:1998ke,Bender:2007nj,Mostafazadeh:2001jk,Balasubramanian:2002zh}. In this process, we discovered that the Hilbert space includes states not observed in Hermitian CFTs, such as those found in \eqref{wwees}. Although examples dealing with non-Hermitian CFTs have been scarce, this paper may provide a starting point for further exploration. However, as this paper deals with non-Hermitian CFTs only formally, more concrete models will need to be developed in order to understand the physical significance of states like those in \eqref{wwees}. One such example is the non-unitary CFT in two dimensions, found in \cite{Hikida:2021ese,Hikida:2022ltr} which has an imaginary-valued central charge. It is obviously important to expand the list of such non-Hermitian CFTs and to try to find any tractable examples among them for our purposes.

{\bf Application to flat spacetimes}

Recently, the application of the holographic principle to asymptotically flat spacetimes, known as celestial holography \cite{Raclariu:2021zjz, Pasterski:2021rjz}, is gaining traction. In the dual CFT of celestial holography, it is known that a special conjugation rule \cite{Pasterski:2021fjn, puhmcelestial, Ogawa:2024nhx}, such as the one used in this study, becomes necessary. It is conceivable that a similar approach could be applied to construct CFTs in celestial holography. However, there are several challenges; one such challenge is the codimension-two nature of the holographic correspondence. Various interpretations have been proposed to address this, such as considering the spacetime as a superposition of (A)dS/CFT by dividing it using (A)dS slices \cite{deBoer:2003vf, Ball:2019atb, Ogawa:2022fhy}. This remains an ongoing work.

\section*{Acknowledgement}

This work is supported by by MEXT KAKENHI Grant-in-Aid for Transformative Research Areas (A) through the ``Extreme Universe'' collaboration: Grant Number 21H05187. TT is also supported by Inamori Research Institute for Science and by JSPS Grant-in-Aid for Scientific Research (A) No.~21H04469. KD is supported by JSPS KAKENHI Grant Number JP24KJ1466. NO is supported by JSPS KAKENHI Grant Number JP24KJ1372. YS is supported by JSPS KAKENHI Grant Number JP23KJ1337.

\appendix

\section{Proof using Pauli matrices}\label{ap:pauli}
In this section, we demonstrate the validity of using Pauli matrices to derive \eqref{global_static_identity}. Consider a set \(\{X_1, X_2, X_3\}\) that follows the $\mathfrak{su}(2)$ commutation relations:
\begin{align}
    [X_i, X_j] = \epsilon_{ijk} X_k.
\end{align}
By applying the Baker-Campbell-Hausdorff (BCH) formula twice, we obtain
\begin{align}\label{BCH_twice}
    e^{X_i} e^{X_j} e^{X_k} = \exp{\sum_{l} a_l X_l}.
\end{align}
This process of applying the BCH formula transforms the exponents into a series involving commutators and their linear combinations:
\begin{align}\label{BCH_formula}
    Z = X + Y + \frac{1}{2}[X, Y] + \frac{1}{12}[X, [X, Y]] - \frac{1}{12}[Y, [X, Y]] + \cdots.
\end{align}
Consequently, the exponent of the right-hand side of (\ref{BCH_twice}) also manifests as a linear combination of \(X_l\). Given that this formulation relies solely on commutators and their linear sums, employing Pauli matrices yields identical outcome. Therefore, Pauli matrices can be used to compute \eqref{global_static_identity}.

However, caution is advised in situations involving products leading to the identity matrix, which falls outside the $\mathfrak{su}(2)$ algebra.

For \eqref{coshid}, Pauli matrices prove to be equally effective. The equation can be expressed using \(\{X_i\}\) as
\begin{align}\label{coshid_alg}
    e^{-2\rho X_1} X_2 e^{2\rho X_1} = X_2 \cosh{2\rho} - 2X_3 \sinh{2\rho}.
\end{align}
The left-hand side (LHS) simplifies through the application of \eqref{BCH_twice} to
\begin{align}
    e^{-2\rho X_1} X_2 e^{2\rho X_1} = \lim_{a \to 0} \frac{d}{da} \left[e^{-2\rho X_1} e^{aX_2} e^{2\rho X_1}\right].
\end{align}
This allows for the LHS of \eqref{coshid_alg} to be computed by applying the BCH formula twice. Initially, we find
\begin{align}
    (\text{LHS}) &= \lim_{a \to 0} \frac{d}{da} \left[e^{f_1(a)X_1 + f_2(a)X_2 + f_3(a)X_3}\right] \notag \\
    &= \left(f'_1(0)X_1 + f'_2(0)X_2 + f'_3(0)X_3\right) e^{f_1(0)X_1 + f_2(0)X_2 + f_3(0)X_3}.
\end{align}
It may initially seem unclear whether Pauli matrices can be used. However, if \(f_1(a)X_1 + f_2(a)X_2 + f_3(a)X_3=O(a)\), the LHS transforms into a linear combination of \(X_i\), confirming the validity of using Pauli matrices.
We can easily prove this as follows.

In evaluating the expression
\begin{align}
e^{-2\rho X_1} e^{aX_2} e^{2\rho X_1},
\end{align}
a detailed inspection of the BCH formula, used twice, reveals that no term is constant with respect to \(a\). Initially, we apply the BCH formula to \(e^{aX_2} e^{2\rho X_1} (:= e^{Z_1})\):
\begin{align}
    Z_1 =& aX_2 + 2\rho X_1 + \frac{1}{2}[aX_2, 2\rho X_1] + \frac{1}{12}[aX_2, [aX_2, 2\rho X_1]] - \frac{1}{12}[Y, [aX_2, 2\rho X_1]] + \cdots \notag \\
    =& 2\rho X_1 + O(a).
\end{align}
Then, we apply it again for \(e^{-2\rho X_1} e^{Z_1} (:= e^{Z_2})\):
\begin{align}
    Z_2 =& -2\rho X_1 + Z_1 + \frac{1}{2}[-2\rho X_1,Z_1] + \frac{1}{12}[-2\rho X_1,[-2\rho X_1,Z_1]] - \frac{1}{12}[Y,[-2\rho X_1,Z_1]] + \cdots\notag\\
    =&O(a).   
\end{align}
On the other hand, since 
\begin{align}
    Z_2=f_1(a)X_1 + f_2(a)X_2 + f_3(a)X_3
\end{align}
holds,
\begin{align}
    f_1(a)X_1 + f_2(a)X_2 + f_3(a)X_3=O(a)
\end{align}
is demonstrated. This assures that the use of Pauli matrices is appropriate, even for \eqref{coshid}.

\section{Calculation of wave function in AdS$_3$}\label{ap:waveads3}
In this appendix, we derive the wave function in AdS$_3$. This provides a detailed account of the calculations presented in \cite{Miyaji:2015fia}. 
\begin{align}
    \Phi(\tau=0,\rho,\phi)&=\bra{\Psi_\a}e^{-\frac{\r}{2}(l_1-l_{-1})}\ket{\a}\nonumber\\
    &=\sum_{k=0}^{\infty}\bra{k}\bra{k}e^{-\frac{i\pi}{2}(L_0+\tilde{L}_0-\Delta_{\alpha})}e^{-\frac{\r}{2}(l_1-l_{-1})}\ket{0}\ket{0}\nonumber\\
    &=\bra{0}e^{\r(L_1-L_{-1})}\ket{0}
\end{align}
We define the following normalization of primary states:
\begin{equation}
    \begin{split}
        &\braket{\a|\a}=1.
    \end{split}
\end{equation}
Then, norm of the descendant states can be calculated as follows:
\begin{align}
    \bra{\a}(L_1)^k(L_{-1})^k\ket{\a}&=  \bra{\a}(L_1)^{k-1}[L_1,(L_1)^k]\ket{\a}\notag\\
    &= \bra{\a}(L_1)^{k-1}\cdot 2(L_{-1})^{k-1}\left(kL_0+\frac{1}{2}k(k-1)\right)\ket{\alpha}\notag\\
    &=\{k\D_\a+k(k-1)\}\bra{\a}(L_1)^{k-1}(L_{-1})^{k-1}\ket{\a}\notag\\
    &=\prod_{l=1}^{k}(l\D_\a+l(l-1)).
\end{align}
So, the normalized descendant states are
\begin{align}
    \ket{k}=\sqrt{\frac{1}{\prod_{l=1}^{k}(l\D_\a+l(l-1))}}(L_{-1})^k\ket{\a},
\end{align}
and we obtain the following recursions
\begin{equation}
    \begin{split}
        &L_1\ket{k}=\sqrt{k\D_\a+k(k-1)}\ket{k-1}\equiv\l_k\ket{k-1},\\
        &L_{-1}\ket{k}=\sqrt{(k+1)\D_\a+k(k+1)}\ket{k+1}\equiv\l_{k+1}\ket{k+1}.
    \end{split}
\end{equation}
Here, we define $f_k(\r)$, $g_k(\r)$, and $h_k(\r)$ as
\begin{equation}\label{fgh}
    \begin{split}
        &f_k(\r)\equiv\bra{k}e^{\r(L_1-L_{-1})}\ket{k},\\
        &g_k(\r)\equiv\bra{k}L_1e^{\r(L_1-L_{-1})}\ket{k}=\l_{k+1}\bra{k+1}e^{\r(L_1-L_{-1})}\ket{k},\\
        &h_k(\r)\equiv\bra{k}L_{-1}e^{\r(L_1-L_{-1})}\ket{k}=\l_{k}\bra{k-1}e^{\r(L_1-L_{-1})}\ket{k}.\\
    \end{split}
\end{equation}
Also note that
\begin{align}\label{gh}
    \bra{k}e^{\r(L_1-L_{-1})}L_1\ket{k}&=\lambda_k  \bra{k}e^{\r(L_1-L_{-1})}\ket{k-1}=h_{k}(-\rho),\nonumber\\
    \bra{k}e^{\r(L_1-L_{-1})}L_{-1}\ket{k}&=\lambda_{k+1}  \bra{k}e^{\r(L_1-L_{-1})}\ket{k+1}=g_k(-\rho).
\end{align}
We can show the following identity from appendix \ref{ap:pauli}.\footnote{
We can replace $L_n~(n=0,\pm1)$ with the following matrix and proceed with the calculations.
\begin{equation}
    \begin{split}
        L_0=\frac{1}{2}
        \left(
        \begin{array}{cc}
            -1 & 0 \\
            0 & 1 \\
        \end{array}
        \right),\quad 
        L_1=
        \left(
        \begin{array}{cc}
            0 & -i \\
            0 & 0 \\
        \end{array}
        \right),\quad    
        L_{-1}=
        \left(
        \begin{array}{cc}
            0 & 0 \\
            -i & 0 \\
        \end{array}
        \right).        
    \end{split}
\end{equation}
}
\begin{align}\label{coshid}
    \bra{k}\left(\frac{L_1+L_{-1}}{2}\right)e^{\r(L_1-L_{-1})}\ket{k}=\bra{k}e^{\r(L_1-L_{-1})}\left[\cosh{2\r}\left(\frac{L_1+L_{-1}}{2}\right)-\sinh{2\r}L_0\right]\ket{k},
\end{align}
Using this result, along with \eqref{fgh} and \eqref{gh}, we obtain the relation
\begin{align}\label{id}
    \frac{1}{2}\left(g_k(\rho)+h_k(\rho)\right)=\frac{1}{2}\cosh{2\rho}\left(g_k(-\rho)+h_k(-\rho)\right)-(\Delta_\alpha+k)\sinh{2\rho}f_k(\rho).
\end{align}
Here we note the following relation:
\begin{align}
    h_{k=0}(\rho)=\bra{0}L_{-1}e^{\r(L_1-L_{-1})}\ket{0}=0.
\end{align}
Differentiating (\ref{fgh}), we get
\begin{align}\label{adsrec2}
    &\frac{\partial}{\partial \r}f_k(\rho)=g_k(\rho)-h_{k}(\rho)=h_k(-\rho)-g_k(-\rho),\nonumber\\
    \Longrightarrow\qquad&~~\qquad\qquad\qquad g_0(\rho)=-g_0(-\rho).
\end{align}
From \eqref{id}, we get the following differential equation,
\begin{align}
    \Delta_{\alpha}\sinh{2\rho}f_0(\rho)&=-\frac{1}{2}(1+\cosh{2\rho})\partial_{\rho}f_0(\rho),\nonumber\\
    \Longrightarrow~\qquad\qquad~\frac{\partial_{\rho}f_0(\rho)}{f_0(\rho)}&=-2\Delta_{\alpha}\frac{\sinh{\rho}}{\cosh{\rho}}
\end{align}
So the wave function in AdS$_3$ can be written as follows:
\begin{align}
    \Phi(\tau=0,\rho,\phi)&=\bra{0}e^{\rho(L_{1}-L_{-1})}\ket{0}=f_0(\rho)\nonumber\\
    &\propto\frac{1}{(\cosh{\rho})^{2\Delta_\alpha}}.
\end{align}
Therefore, 
\begin{align}
    \Phi(\tau,\rho,\phi)&=\bra{\Psi(\rho, \phi)}e^{i\tau(L_0+\tilde{L}_0)}\ket{\alpha}\propto\frac{e^{-2it\Delta_{\alpha}\tau}}{(\cosh{\rho})^{2\Delta_\alpha}}.
\end{align}
Using this result, $\Phi_{k,k}(\tau,\rho,\phi)$ can also be derived. 
For more details, refer to \cite{Miyaji:2015fia}.

\section{Scalar wave function}\label{ap:scalar}

Here we analyze scalar wave functions in static dS$_3$.
As we saw previously in (\ref{wave}), the wave function for a free massive scalar $\Phi_{k,k}$ is given by
\be
\Phi^{\rm (AdS)}_{k,\bar{k}}(\tau,\rho,x)=e^{-i(\Delta +k+\bar{k})\tau}e^{i(k-\bar{k})x}
(\tanh\rho)^{|k-\bar{k}|}(\cosh\rho)^{-\Delta}\cdot P^{(k-\bar{k},\Delta-1)}_{\bar{k}}\left(\frac{2}{\cosh^2\rho}-1\right),\nn
\ee
Via the analytic continuation to  static patch dS$_3$, the scalar field takes the form
\ba
\Phi^{(dS)}_{k,\bar{k}}(t,\theta,\phi)=(i)^{|k-\bar{k}|}e^{(\Delta +k+\bar{k})t}e^{i(k-\bar{k})\phi}
(\tan\theta)^{|k-\bar{k}|}(\cos\theta)^{-\Delta}\cdot P^{(k-\bar{k},\Delta-1)}_{\bar{k}}\left(\frac{2}{\cos^2\theta}-1\right).\no
\ea
We can rewrite the $P$-function as follows
\ba
P^{(k-\bar{k},\Delta-1)}_{\bar{k}}\left(\frac{2}{\cos^2\theta}-1\right)
&=&{}_2F_1\left[-\bar{k},\bar{k}+|k-\bar{k}|+\Delta;|k-\bar{k}|+1;-\tan^2\theta\right]\no
&=&(\cos\theta)^{-\bar{k}}\cdot {}_2F_1\left[-\bar{k},1-\bar{k}-\Delta;|k-\bar{k}|+1;\sin^2\theta\right].
\ea
Thus we obtain
\ba
\Phi^{(dS)}_{k,\bar{k}}(t,\theta,\phi)&=&(i)^{|k-\bar{k}|}e^{(\Delta +k+\bar{k})t}e^{i(k-\bar{k})\phi}
(\sin\theta)^{|k-\bar{k}|}(\cos\theta)^{-\Delta-2\bar{k}-|k-\bar{k}|}\no
&& \times
 {}_2F_1\left[-\bar{k},1-\bar{k}-\Delta;|k-\bar{k}|+1;\sin^2\theta\right].\label{dssca}
\ea

On the other hand, according to \cite{Bousso:2001mw}, the scalar field solution on dS$_3$ that is regular near the south pole reads 
\ba
&& \Phi^{\rm (S)}(t,\theta,\phi) \no
&& =e^{-i\omega t+iJ\phi}(\sin\theta)^{|J|}(\cos\theta)^{i\omega}
\times  {}_2F_1\left[\frac{|J|+i\omega+\Delta}{2},\frac{|J|+i\omega+2-\Delta}{2}
;1+|J|;\sin^2\theta\right].\no  \label{dsscb}
\ea

By comparing (\ref{dssca}) with (\ref{dsscb}), we find that they coincide by identifying the parameters as
\ba
J=k-\bar{k},\ \ \ \omega=i(\Delta+k+\bar{k}).
\ea

This identification is what we expected since we know the evolution of $t$ and $\phi$ directions in static dS$_3$ is given by $e^{t(L_0+\bar{L}_0)}e^{-i\phi(L_0-\bar{L}_0)}$.

In the dual CFT, $k$ and $\bar{k}$ are levels of Virasoro descendants and they should be integer-valued.
It is intriguing to note that if we require the ingoing boundary condition at the horizon, we find $k$ and $\bar{k}$ are non-negative integers. From the CFT side, this is natural since $k$ and $\bar{k}$ correspond to the numbers of descendants acting on primary states. Indeed assuming $k>\bar{k}$, the Gamma function $\Gamma\left(\frac{1}{2}(|J|+i\omega+\Delta)\right)=\Gamma(-\bar{k})$ in (6.9) of \cite{Bousso:2001mw} becomes divergent for $\bar{k}=0,1,2,\ddd$. This might explain the quantization of $(k,\bar{k})$.

\section{de Sitter CFT as a non-Hermitian system}\label{appendix_nonhemitian}
\subsubsection*{Brief review of non-Hermitian systems}
In non-Hermitian quantum mechanics, where \(H \neq H^\dagger\), we must revise the rules for conjugation to ensure a well-defined density matrix for pure states. We define the density matrix \(\rho\) for state \(\ket{E_n}\) as follows
\begin{align}
    \rho := \ket{E_n}\bra{E_n},
\end{align}
where the eigenvalue equation for \(\ket{E_n}\) and its dual \(\bra{E_n}\) are given by
\begin{align}
    H\ket{E_n} = E_n\ket{E_n}, \qquad \bra{E_n}H = E_n\bra{E_n}.
\end{align}
The bra state \(\bra{E_n}\) satisfies the relations
\begin{align}
   \bra{E_n} = \ket{\widehat{E_n}}^\dagger, \qquad H^\dagger\ket{\widehat{E_n}} = E_n^*\ket{\widehat{E_n}}.
\end{align}
Consequently, we obtain the following trace conditions:
\begin{align}
    \Tr(\rho H) = \Tr(H\rho) = E_n.
\end{align}
Thus, the conjugate is defined as
\begin{align}\label{nonHermitian_normalization}
    \ket{E_n} \to \bra{E_n} = \ket{\widehat{E_n}}^\dagger.
\end{align}
Assuming no degeneracy among the eigenvectors, we have:
\begin{align}\label{normalization}
    \braket{E_m|E_n} = \delta_{mn}.
\end{align}
This is justified because 
\begin{align}
    \bra{E_m}H\ket{E_n} = E_m\braket{E_m|E_n} = E_n\braket{E_m|E_n}.
\end{align}
In contrast, the Hermitian conjugate of \(\ket{E_n}\) satisfies\footnote{Note that \(\braket{\widehat{E_m}|E_n} \neq \delta_{mn}\), but $\braket{\widehat{E_m}|E_n}$ is usually finite.}
\begin{align}
    \bra{\widehat{E_n}} := (\ket{E_n})^\dagger, \qquad \bra{\widehat{E_n}}H^\dagger = E_n^*\bra{\widehat{E_n}}.
\end{align}
Of course, the following is also valid:
\begin{align}\label{normalization_hat}
    \braket{\widehat{E_m}|\widehat{E_n}} = \delta_{mn}.
\end{align}

\subsubsection*{Application to our system}
In the original  dS$_3$ conjugation, the operators satisfy
\begin{align}
    L_0^\dagger = -\ti{L}_0, \quad L_{\pm 1}^\dagger = \tilde{L}_{\pm 1}.
\end{align}
Considering a primary state \( \ket{\Delta} \), which possesses a conformal weight defined by \( L_0 \ket{\Delta} =  \ti{L}_0 \ket{\Delta} =\Delta \ket{\Delta} \).
Thus its Hermitian conjugate \( \bra{\widehat{\Delta}} \) satisfies
\begin{align}
    \bra{\widehat{\Delta}} L_0 = -\frac{\Delta^*}{2} \bra{\widehat{\Delta}}.
\end{align}
This leads to the equation
\begin{align}
    \braket{\widehat{\Delta_i}|L_0|\Delta_j} = \frac{\Delta_j}{2} \braket{\widehat{\Delta_i}|\Delta_j} = -\frac{\Delta_i^*}{2} \braket{\widehat{\Delta_i}|\Delta_j}.
\end{align}
Therefore, we get the following:
\begin{align}
    -\Delta_i^* = \Delta_j \quad \text{or} \quad \braket{\widehat{\Delta_i}|\Delta_j} = 0.
\end{align}
In our de Sitter scenario, assuming mass \( m > 1 \), each eigenvalue is expressed as \( \Delta = 1 \pm i\mu \), leading to
\begin{align}\label{orthogonal}
    \braket{\widehat{\Delta_i}|\Delta_j} = 0 \quad \text{for} \ \forall \Delta.
\end{align}
On the other hand, we define bra states $\bra{\Delta}$ as
\begin{align}
   \bra{\Delta} = (\ket{\widehat{\Delta}})^\dagger, \qquad L_0^\dagger\ket{\widehat{\Delta}} =\frac{\Delta^*}{2}\ket{\widehat{\Delta}}.
\end{align}
We can normalize these states in a similar way as \eqref{nonHermitian_normalization} and \eqref{normalization_hat}:
\begin{align}\label{normalization_dS}
    \begin{split}
        &\braket{\Delta_i|\Delta_j}=\braket{\widehat{\Delta_i}| \widehat{\Delta_j}}=\delta_{ij}.
    \end{split}
\end{align}
By combining \eqref{orthogonal} and \eqref{normalization_dS}, we can demonstrate \eqref{normalization_dS_3}.

\section{Some useful identities}\label{ap:formula}

For the $\mathfrak{su}(2)$ algebra $(J_1,J_2,J_3)$  defined in (\ref{jsut}), we can easily show
\ba
&& e^{i\beta J_1}J_2 e^{-i\beta J_1}=\cos\beta J_2-\sin\beta J_3,\ \ \  \ e^{i\beta J_1}J_3 e^{-i\beta J_1}=\cos\beta J_3+\sin\beta J_2,\no
&& e^{i\beta J_2}J_3 e^{-i\beta J_2}=\cos\beta J_3-\sin\beta J_1,\ \ \ \  e^{i\beta J_2}J_1 e^{-i\beta J_2}=\cos\beta J_1+\sin\beta J_3,\no
&&e^{i\beta J_3}J_1 e^{-i\beta J_3}=\cos\beta J_1-\sin\beta J_2,\ \ \ \  e^{i\beta J_3}J_2 e^{-i\beta J_3}=\cos\beta J_2+\sin\beta J_1.
\label{wwwq}
\ea
Note that the left-moving and right-moving sector are completely decoupled in the above identities. Therefore we can regard 
\ba
J_{1}\to \frac{1}{2}(L_1-L_{-1}),\ \ \ J_2\to\frac{i}{2}(L_1+L_{-1}),\ \ \ 
J_3\to L_0,
\ea
to find
\ba
&& e^{i\beta J_1}L_0 e^{-i\beta J_1}=\cos\beta L_0+\frac{i}{2}\sin\beta (L_1+L_{-1}),\no
&& e^{i\beta J_1}L_{\pm 1} e^{-i\beta J_1}=\pm\frac{1}{2}(L_1-L_{-1})+\frac{1}{2}\cos\beta(L_1+L_{-1})+i\sin\beta L_0,\no
&& e^{i\beta J_3}L_{\pm 1} e^{-i\beta J_3}=e^{\mp i\beta}L_{\pm 1},
\ea
which leads to (\ref{Jpi}).

\clearpage
\bibliographystyle{JHEP}
\bibliography{dSCFT}

\providecommand{\href}[2]{#2}\begingroup\raggedright\begin{thebibliography}{10}

\bibitem{tHooft:1993dmi}
G.~'t~Hooft, \emph{{Dimensional reduction in quantum gravity}}, {\emph{Conf. Proc. C} {\bfseries 930308} (1993) 284} [\href{https://arxiv.org/abs/gr-qc/9310026}{{\ttfamily gr-qc/9310026}}].

\bibitem{Susskind:1994vu}
L.~Susskind, \emph{{The World as a hologram}}, \href{https://doi.org/10.1063/1.531249}{\emph{J. Math. Phys.} {\bfseries 36} (1995) 6377} [\href{https://arxiv.org/abs/hep-th/9409089}{{\ttfamily hep-th/9409089}}].

\bibitem{Maldacena:1997re}
J.~M. Maldacena, \emph{{The Large N limit of superconformal field theories and supergravity}}, \href{https://doi.org/10.1023/A:1026654312961}{\emph{Adv. Theor. Math. Phys.} {\bfseries 2} (1998) 231} [\href{https://arxiv.org/abs/hep-th/9711200}{{\ttfamily hep-th/9711200}}].

\bibitem{Gubser:1998bc}
S.~S. Gubser, I.~R. Klebanov and A.~M. Polyakov, \emph{{Gauge theory correlators from noncritical string theory}}, \href{https://doi.org/10.1016/S0370-2693(98)00377-3}{\emph{Phys. Lett. B} {\bfseries 428} (1998) 105} [\href{https://arxiv.org/abs/hep-th/9802109}{{\ttfamily hep-th/9802109}}].

\bibitem{Witten:1998qj}
E.~Witten, \emph{{Anti-de Sitter space and holography}}, \href{https://doi.org/10.4310/ATMP.1998.v2.n2.a2}{\emph{Adv. Theor. Math. Phys.} {\bfseries 2} (1998) 253} [\href{https://arxiv.org/abs/hep-th/9802150}{{\ttfamily hep-th/9802150}}].

\bibitem{Strominger:2001pn}
A.~Strominger, \emph{{The dS / CFT correspondence}}, \href{https://doi.org/10.1088/1126-6708/2001/10/034}{\emph{JHEP} {\bfseries 10} (2001) 034} [\href{https://arxiv.org/abs/hep-th/0106113}{{\ttfamily hep-th/0106113}}].

\bibitem{Witten:2001kn}
E.~Witten, \emph{{Quantum gravity in de Sitter space}},  in \emph{{Strings 2001: International Conference}}, 6, 2001, \href{https://arxiv.org/abs/hep-th/0106109}{{\ttfamily hep-th/0106109}}.

\bibitem{Maldacena:2002vr}
J.~M. Maldacena, \emph{{Non-Gaussian features of primordial fluctuations in single field inflationary models}}, \href{https://doi.org/10.1088/1126-6708/2003/05/013}{\emph{JHEP} {\bfseries 05} (2003) 013} [\href{https://arxiv.org/abs/astro-ph/0210603}{{\ttfamily astro-ph/0210603}}].

\bibitem{Bousso:2001mw}
R.~Bousso, A.~Maloney and A.~Strominger, \emph{{Conformal vacua and entropy in de Sitter space}}, \href{https://doi.org/10.1103/PhysRevD.65.104039}{\emph{Phys. Rev. D} {\bfseries 65} (2002) 104039} [\href{https://arxiv.org/abs/hep-th/0112218}{{\ttfamily hep-th/0112218}}].

\bibitem{Spradlin:2001nb}
M.~Spradlin and A.~Volovich, \emph{{Vacuum states and the S matrix in dS / CFT}}, \href{https://doi.org/10.1103/PhysRevD.65.104037}{\emph{Phys. Rev. D} {\bfseries 65} (2002) 104037} [\href{https://arxiv.org/abs/hep-th/0112223}{{\ttfamily hep-th/0112223}}].

\bibitem{Guijosa:2003ze}
A.~Guijosa and D.~A. Lowe, \emph{{A New twist on dS / CFT}}, \href{https://doi.org/10.1103/PhysRevD.69.106008}{\emph{Phys. Rev. D} {\bfseries 69} (2004) 106008} [\href{https://arxiv.org/abs/hep-th/0312282}{{\ttfamily hep-th/0312282}}].

\bibitem{Ness:2005qj}
S.~Ness and G.~Siopsis, \emph{{Virasoro generators and the dS(3)/CFT(2) correspondence}}, \href{https://doi.org/10.1088/0264-9381/22/17/027}{\emph{Class. Quant. Grav.} {\bfseries 22} (2005) 3789} [\href{https://arxiv.org/abs/hep-th/0505146}{{\ttfamily hep-th/0505146}}].

\bibitem{Harlow:2011ke}
D.~Harlow and D.~Stanford, \emph{{Operator Dictionaries and Wave Functions in AdS/CFT and dS/CFT}},  \href{https://arxiv.org/abs/1104.2621}{{\ttfamily 1104.2621}}.

\bibitem{Castro:2023bvo}
A.~Castro, I.~Coman, J.~R. Fliss and C.~Zukowski, \emph{{Coupling Fields to 3D Quantum Gravity via Chern-Simons Theory}}, \href{https://doi.org/10.1103/PhysRevLett.131.171602}{\emph{Phys. Rev. Lett.} {\bfseries 131} (2023) 171602} [\href{https://arxiv.org/abs/2304.02668}{{\ttfamily 2304.02668}}].

\bibitem{Miyaji:2015fia}
M.~Miyaji, T.~Numasawa, N.~Shiba, T.~Takayanagi and K.~Watanabe, \emph{{Continuous Multiscale Entanglement Renormalization Ansatz as Holographic Surface-State Correspondence}}, \href{https://doi.org/10.1103/PhysRevLett.115.171602}{\emph{Phys. Rev. Lett.} {\bfseries 115} (2015) 171602} [\href{https://arxiv.org/abs/1506.01353}{{\ttfamily 1506.01353}}].

\bibitem{Goto:2016wme}
K.~Goto, M.~Miyaji and T.~Takayanagi, \emph{{Causal Evolutions of Bulk Local Excitations from CFT}}, \href{https://doi.org/10.1007/JHEP09(2016)130}{\emph{JHEP} {\bfseries 09} (2016) 130} [\href{https://arxiv.org/abs/1605.02835}{{\ttfamily 1605.02835}}].

\bibitem{Goto:2017olq}
K.~Goto and T.~Takayanagi, \emph{{CFT descriptions of bulk local states in the AdS black holes}}, \href{https://doi.org/10.1007/JHEP10(2017)153}{\emph{JHEP} {\bfseries 10} (2017) 153} [\href{https://arxiv.org/abs/1704.00053}{{\ttfamily 1704.00053}}].

\bibitem{Nakayama:2015mva}
Y.~Nakayama and H.~Ooguri, \emph{{Bulk Locality and Boundary Creating Operators}}, \href{https://doi.org/10.1007/JHEP10(2015)114}{\emph{JHEP} {\bfseries 10} (2015) 114} [\href{https://arxiv.org/abs/1507.04130}{{\ttfamily 1507.04130}}].

\bibitem{Verlinde:2015qfa}
H.~Verlinde, \emph{{Poking Holes in AdS/CFT: Bulk Fields from Boundary States}},  \href{https://arxiv.org/abs/1505.05069}{{\ttfamily 1505.05069}}.

\bibitem{Lewkowycz:2016ukf}
A.~Lewkowycz, G.~J. Turiaci and H.~Verlinde, \emph{{A CFT Perspective on Gravitational Dressing and Bulk Locality}}, \href{https://doi.org/10.1007/JHEP01(2017)004}{\emph{JHEP} {\bfseries 01} (2017) 004} [\href{https://arxiv.org/abs/1608.08977}{{\ttfamily 1608.08977}}].

\bibitem{Hamilton:2005ju}
A.~Hamilton, D.~N. Kabat, G.~Lifschytz and D.~A. Lowe, \emph{{Local bulk operators in AdS/CFT: A Boundary view of horizons and locality}}, \href{https://doi.org/10.1103/PhysRevD.73.086003}{\emph{Phys. Rev. D} {\bfseries 73} (2006) 086003} [\href{https://arxiv.org/abs/hep-th/0506118}{{\ttfamily hep-th/0506118}}].

\bibitem{Hamilton:2006az}
A.~Hamilton, D.~N. Kabat, G.~Lifschytz and D.~A. Lowe, \emph{{Holographic representation of local bulk operators}}, \href{https://doi.org/10.1103/PhysRevD.74.066009}{\emph{Phys. Rev. D} {\bfseries 74} (2006) 066009} [\href{https://arxiv.org/abs/hep-th/0606141}{{\ttfamily hep-th/0606141}}].

\bibitem{Xiao:2014uea}
X.~Xiao, \emph{{Holographic representation of local operators in de sitter space}}, \href{https://doi.org/10.1103/PhysRevD.90.024061}{\emph{Phys. Rev. D} {\bfseries 90} (2014) 024061} [\href{https://arxiv.org/abs/1402.7080}{{\ttfamily 1402.7080}}].

\bibitem{Chatterjee:2015pha}
A.~Chatterjee and D.~A. Lowe, \emph{{Holographic operator mapping in dS/CFT and cluster decomposition}}, \href{https://doi.org/10.1103/PhysRevD.92.084038}{\emph{Phys. Rev. D} {\bfseries 92} (2015) 084038} [\href{https://arxiv.org/abs/1503.07482}{{\ttfamily 1503.07482}}].

\bibitem{Chatterjee:2016ifv}
A.~Chatterjee and D.~A. Lowe, \emph{{dS/CFT and the operator product expansion}}, \href{https://doi.org/10.1103/PhysRevD.96.066031}{\emph{Phys. Rev. D} {\bfseries 96} (2017) 066031} [\href{https://arxiv.org/abs/1612.07785}{{\ttfamily 1612.07785}}].

\bibitem{Bhowmick:2019nso}
S.~Bhowmick, K.~Ray and S.~Sen, \emph{{Holography in de Sitter and anti-de Sitter Spaces and Gel'fand Graev Radon transform}}, \href{https://doi.org/10.1016/j.physletb.2019.134977}{\emph{Phys. Lett. B} {\bfseries 798} (2019) 134977} [\href{https://arxiv.org/abs/1903.07336}{{\ttfamily 1903.07336}}].

\bibitem{Sleight:2020obc}
C.~Sleight and M.~Taronna, \emph{{From AdS to dS exchanges: Spectral representation, Mellin amplitudes, and crossing}}, \href{https://doi.org/10.1103/PhysRevD.104.L081902}{\emph{Phys. Rev. D} {\bfseries 104} (2021) L081902} [\href{https://arxiv.org/abs/2007.09993}{{\ttfamily 2007.09993}}].

\bibitem{Sleight:2021plv}
C.~Sleight and M.~Taronna, \emph{{From dS to AdS and back}}, \href{https://doi.org/10.1007/JHEP12(2021)074}{\emph{JHEP} {\bfseries 12} (2021) 074} [\href{https://arxiv.org/abs/2109.02725}{{\ttfamily 2109.02725}}].

\bibitem{Anninos:2011ui}
D.~Anninos, T.~Hartman and A.~Strominger, \emph{{Higher Spin Realization of the dS/CFT Correspondence}}, \href{https://doi.org/10.1088/1361-6382/34/1/015009}{\emph{Class. Quant. Grav.} {\bfseries 34} (2017) 015009} [\href{https://arxiv.org/abs/1108.5735}{{\ttfamily 1108.5735}}].

\bibitem{Ng:2012xp}
G.~S. Ng and A.~Strominger, \emph{{State/Operator Correspondence in Higher-Spin dS/CFT}}, \href{https://doi.org/10.1088/0264-9381/30/10/104002}{\emph{Class. Quant. Grav.} {\bfseries 30} (2013) 104002} [\href{https://arxiv.org/abs/1204.1057}{{\ttfamily 1204.1057}}].

\bibitem{Hikida:2021ese}
Y.~Hikida, T.~Nishioka, T.~Takayanagi and Y.~Taki, \emph{{Holography in de Sitter Space via Chern-Simons Gauge Theory}}, \href{https://doi.org/10.1103/PhysRevLett.129.041601}{\emph{Phys. Rev. Lett.} {\bfseries 129} (2022) 041601} [\href{https://arxiv.org/abs/2110.03197}{{\ttfamily 2110.03197}}].

\bibitem{Hikida:2022ltr}
Y.~Hikida, T.~Nishioka, T.~Takayanagi and Y.~Taki, \emph{{CFT duals of three-dimensional de Sitter gravity}}, \href{https://doi.org/10.1007/JHEP05(2022)129}{\emph{JHEP} {\bfseries 05} (2022) 129} [\href{https://arxiv.org/abs/2203.02852}{{\ttfamily 2203.02852}}].

\bibitem{Alishahiha:2004md}
M.~Alishahiha, A.~Karch, E.~Silverstein and D.~Tong, \emph{{The dS/dS correspondence}}, \href{https://doi.org/10.1063/1.1848341}{\emph{AIP Conf. Proc.} {\bfseries 743} (2004) 393} [\href{https://arxiv.org/abs/hep-th/0407125}{{\ttfamily hep-th/0407125}}].

\bibitem{Dong:2018cuv}
X.~Dong, E.~Silverstein and G.~Torroba, \emph{{De Sitter holography and entanglement entropy}}, \href{https://doi.org/10.1007/JHEP07(2018)050}{\emph{JHEP} {\bfseries 07} (2018) 050} [\href{https://arxiv.org/abs/1804.08623}{{\ttfamily 1804.08623}}].

\bibitem{Gorbenko:2018oov}
V.~Gorbenko, E.~Silverstein and G.~Torroba, \emph{{dS/dS and $ T\overline{T} $}}, \href{https://doi.org/10.1007/JHEP03(2019)085}{\emph{JHEP} {\bfseries 03} (2019) 085} [\href{https://arxiv.org/abs/1811.07965}{{\ttfamily 1811.07965}}].

\bibitem{Narayan:2015vda}
K.~Narayan, \emph{{Extremal surfaces in de Sitter spacetime}}, \href{https://doi.org/10.1103/PhysRevD.91.126011}{\emph{Phys. Rev. D} {\bfseries 91} (2015) 126011} [\href{https://arxiv.org/abs/1501.03019}{{\ttfamily 1501.03019}}].

\bibitem{Sato:2015tta}
Y.~Sato, \emph{{Comments on entanglement entropy in the dS/CFT correspondence}}, \href{https://doi.org/10.1103/PhysRevD.91.086009}{\emph{Phys. Rev. D} {\bfseries 91} (2015) 086009} [\href{https://arxiv.org/abs/1501.04903}{{\ttfamily 1501.04903}}].

\bibitem{Miyaji:2015yva}
M.~Miyaji and T.~Takayanagi, \emph{{Surface/State Correspondence as a Generalized Holography}}, \href{https://doi.org/10.1093/ptep/ptv089}{\emph{PTEP} {\bfseries 2015} (2015) 073B03} [\href{https://arxiv.org/abs/1503.03542}{{\ttfamily 1503.03542}}].

\bibitem{Geng:2021wcq}
H.~Geng, Y.~Nomura and H.-Y. Sun, \emph{{Information paradox and its resolution in de Sitter holography}}, \href{https://doi.org/10.1103/PhysRevD.103.126004}{\emph{Phys. Rev. D} {\bfseries 103} (2021) 126004} [\href{https://arxiv.org/abs/2103.07477}{{\ttfamily 2103.07477}}].

\bibitem{Doi:2022iyj}
K.~Doi, J.~Harper, A.~Mollabashi, T.~Takayanagi and Y.~Taki, \emph{{Pseudoentropy in dS/CFT and Timelike Entanglement Entropy}}, \href{https://doi.org/10.1103/PhysRevLett.130.031601}{\emph{Phys. Rev. Lett.} {\bfseries 130} (2023) 031601} [\href{https://arxiv.org/abs/2210.09457}{{\ttfamily 2210.09457}}].

\bibitem{Cotler:2023xku}
J.~Cotler and A.~Strominger, \emph{{Cosmic ER=EPR in dS/CFT}},  \href{https://arxiv.org/abs/2302.00632}{{\ttfamily 2302.00632}}.

\bibitem{Kawamoto:2023nki}
T.~Kawamoto, S.-M. Ruan, Y.-k. Suzuki and T.~Takayanagi, \emph{{A Half de Sitter Holography}},  \href{https://arxiv.org/abs/2306.07575}{{\ttfamily 2306.07575}}.

\bibitem{Susskind:2021esx}
L.~Susskind, \emph{{Entanglement and chaos in de Sitter holography: An SYK example}},  \href{https://arxiv.org/abs/2109.14104}{{\ttfamily 2109.14104}}.

\bibitem{Chandrasekaran:2022cip}
V.~Chandrasekaran, R.~Longo, G.~Penington and E.~Witten, \emph{{An algebra of observables for de Sitter space}}, \href{https://doi.org/10.1007/JHEP02(2023)082}{\emph{JHEP} {\bfseries 02} (2023) 082} [\href{https://arxiv.org/abs/2206.10780}{{\ttfamily 2206.10780}}].

\bibitem{Maldacena:2019cbz}
J.~Maldacena, G.~J. Turiaci and Z.~Yang, \emph{{Two dimensional Nearly de Sitter gravity}}, \href{https://doi.org/10.1007/JHEP01(2021)139}{\emph{JHEP} {\bfseries 01} (2021) 139} [\href{https://arxiv.org/abs/1904.01911}{{\ttfamily 1904.01911}}].

\bibitem{Cotler:2019nbi}
J.~Cotler, K.~Jensen and A.~Maloney, \emph{{Low-dimensional de Sitter quantum gravity}}, \href{https://doi.org/10.1007/JHEP06(2020)048}{\emph{JHEP} {\bfseries 06} (2020) 048} [\href{https://arxiv.org/abs/1905.03780}{{\ttfamily 1905.03780}}].

\bibitem{Susskind:2022bia}
L.~Susskind, \emph{{De Sitter Space, Double-Scaled SYK, and the Separation of Scales in the Semiclassical Limit}},  \href{https://arxiv.org/abs/2209.09999}{{\ttfamily 2209.09999}}.

\bibitem{Narovlansky:2023lfz}
V.~Narovlansky and H.~Verlinde, \emph{{Double-scaled SYK and de Sitter Holography}},  \href{https://arxiv.org/abs/2310.16994}{{\ttfamily 2310.16994}}.

\bibitem{Bunch:1978yq}
T.~S. Bunch and P.~C.~W. Davies, \emph{{Quantum Field Theory in de Sitter Space: Renormalization by Point Splitting}}, \href{https://doi.org/10.1098/rspa.1978.0060}{\emph{Proc. Roy. Soc. Lond. A} {\bfseries 360} (1978) 117}.

\bibitem{Hartle:1983ai}
J.~B. Hartle and S.~W. Hawking, \emph{{Wave Function of the Universe}}, \href{https://doi.org/10.1103/PhysRevD.28.2960}{\emph{Phys. Rev. D} {\bfseries 28} (1983) 2960}.

\bibitem{Harlow:2023hjb}
D.~Harlow and T.~Numasawa, \emph{{Gauging spacetime inversions in quantum gravity}},  \href{https://arxiv.org/abs/2311.09978}{{\ttfamily 2311.09978}}.

\bibitem{Susskind:2023rxm}
L.~Susskind, \emph{{A Paradox and its Resolution Illustrate Principles of de Sitter Holography}},  \href{https://arxiv.org/abs/2304.00589}{{\ttfamily 2304.00589}}.

\bibitem{Goodhew:2024eup}
H.~Goodhew, A.~Thavanesan and A.~C. Wall, \emph{{The Cosmological CPT Theorem}},  \href{https://arxiv.org/abs/2408.17406}{{\ttfamily 2408.17406}}.

\bibitem{Nakata:2020luh}
Y.~Nakata, T.~Takayanagi, Y.~Taki, K.~Tamaoka and Z.~Wei, \emph{{New holographic generalization of entanglement entropy}}, \href{https://doi.org/10.1103/PhysRevD.103.026005}{\emph{Phys. Rev. D} {\bfseries 103} (2021) 026005} [\href{https://arxiv.org/abs/2005.13801}{{\ttfamily 2005.13801}}].

\bibitem{Bender:1998ke}
C.~M. Bender and S.~Boettcher, \emph{{Real spectra in nonHermitian Hamiltonians having PT symmetry}}, \href{https://doi.org/10.1103/PhysRevLett.80.5243}{\emph{Phys. Rev. Lett.} {\bfseries 80} (1998) 5243} [\href{https://arxiv.org/abs/physics/9712001}{{\ttfamily physics/9712001}}].

\bibitem{Bender:2007nj}
C.~M. Bender, \emph{{Making sense of non-Hermitian Hamiltonians}}, \href{https://doi.org/10.1088/0034-4885/70/6/R03}{\emph{Rept. Prog. Phys.} {\bfseries 70} (2007) 947} [\href{https://arxiv.org/abs/hep-th/0703096}{{\ttfamily hep-th/0703096}}].

\bibitem{Mostafazadeh:2001jk}
A.~Mostafazadeh, \emph{{PseudoHermiticity versus PT symmetry. The necessary condition for the reality of the spectrum}}, \href{https://doi.org/10.1063/1.1418246}{\emph{J. Math. Phys.} {\bfseries 43} (2002) 205} [\href{https://arxiv.org/abs/math-ph/0107001}{{\ttfamily math-ph/0107001}}].

\bibitem{Balasubramanian:2002zh}
V.~Balasubramanian, J.~de~Boer and D.~Minic, \emph{{Notes on de Sitter space and holography}}, \href{https://doi.org/10.1016/S0003-4916(02)00020-9}{\emph{Class. Quant. Grav.} {\bfseries 19} (2002) 5655} [\href{https://arxiv.org/abs/hep-th/0207245}{{\ttfamily hep-th/0207245}}].

\bibitem{Raclariu:2021zjz}
A.-M. Raclariu, \emph{{Lectures on Celestial Holography}},  \href{https://arxiv.org/abs/2107.02075}{{\ttfamily 2107.02075}}.

\bibitem{Pasterski:2021rjz}
S.~Pasterski, \emph{{Lectures on celestial amplitudes}}, \href{https://doi.org/10.1140/epjc/s10052-021-09846-7}{\emph{Eur. Phys. J. C} {\bfseries 81} (2021) 1062} [\href{https://arxiv.org/abs/2108.04801}{{\ttfamily 2108.04801}}].

\bibitem{Pasterski:2021fjn}
S.~Pasterski, A.~Puhm and E.~Trevisani, \emph{{Celestial diamonds: conformal multiplets in celestial CFT}}, \href{https://doi.org/10.1007/JHEP11(2021)072}{\emph{JHEP} {\bfseries 11} (2021) 072} [\href{https://arxiv.org/abs/2105.03516}{{\ttfamily 2105.03516}}].

\bibitem{puhmcelestial}
A.~Puhm, \emph{A celestial holography primer},  2021.

\bibitem{Ogawa:2024nhx}
N.~Ogawa, S.~Takahashi, T.~Tsuda and T.~Waki, \emph{{Celestial CFT from $H_3^+$-WZW Model}},  \href{https://arxiv.org/abs/2404.12049}{{\ttfamily 2404.12049}}.

\bibitem{deBoer:2003vf}
J.~de~Boer and S.~N. Solodukhin, \emph{{A Holographic reduction of Minkowski space-time}}, \href{https://doi.org/10.1016/S0550-3213(03)00494-2}{\emph{Nucl. Phys. B} {\bfseries 665} (2003) 545} [\href{https://arxiv.org/abs/hep-th/0303006}{{\ttfamily hep-th/0303006}}].

\bibitem{Ball:2019atb}
A.~Ball, E.~Himwich, S.~A. Narayanan, S.~Pasterski and A.~Strominger, \emph{{Uplifting AdS$_{3}$/CFT$_{2}$ to flat space holography}}, \href{https://doi.org/10.1007/JHEP08(2019)168}{\emph{JHEP} {\bfseries 08} (2019) 168} [\href{https://arxiv.org/abs/1905.09809}{{\ttfamily 1905.09809}}].

\bibitem{Ogawa:2022fhy}
N.~Ogawa, T.~Takayanagi, T.~Tsuda and T.~Waki, \emph{{Wedge holography in flat space and celestial holography}}, \href{https://doi.org/10.1103/PhysRevD.107.026001}{\emph{Phys. Rev. D} {\bfseries 107} (2023) 026001} [\href{https://arxiv.org/abs/2207.06735}{{\ttfamily 2207.06735}}].

\end{thebibliography}\endgroup


\end{document}